\documentclass{aa}

\usepackage{graphicx}
\usepackage{txfonts}
\usepackage{lscape}
\usepackage{hyperref}

\usepackage{xcolor}
\graphicspath{Figures/}
\newcommand{\cmfast}{{\tt 21cmFAST}}
\newcommand{\cmfastvfour}{{\tt 21cmFASTv4}}
\newcommand{\gcheck}{\textcolor{green}{\checkmark}}
\newcommand{\rcross}{\textcolor{red}{\text{\sffamily X}}}
\newenvironment{packed_enum}{
\begin{enumerate}[(i)]
  \setlength{\itemsep}{1pt}
  \setlength{\parskip}{0pt}
  \setlength{\parsep}{0pt}
}{\end{enumerate}}

\begin{document}

   \title{Efficient simulation of discrete galaxy populations and associated radiation fields during the first billion years}
   \titlerunning{21cmFASTv4: Discrete Galaxy Lightcones}


   \author{James E. Davies\thanks{email:james.davies@sns.it}
          \inst{1}
          \and
          Andrei Mesinger\inst{1, 2}
          \and
          Steven Murray\inst{1}}

   \institute{Scuola Normale Superiore di Pisa,
              Piazza dei Cavallieri 7, 56126 Pisa, Italy
             \and 
            Centro Nazionale ''High Performance Computing, Big Data and Quantum Computing''\\
            }

   \date{Received XXXX; accepted XXXX}

  \abstract
    {Understanding the epochs of cosmic dawn and reionisation requires us to leverage multi-wavelength and multi-tracer observations, with each dataset providing a complimentary piece of the puzzle.
    To interpret such data, we update the public simulation code, \cmfastvfour, to include a  discrete source model based on stochastic sampling of conditional mass functions and semi-empirical galaxy relations. 
We demonstrate that our new galaxy model, which parametrizes the means and scatters of well-established scaling relations, is flexible enough to characterize very different predictions from hydrodynamic cosmological simulations of high-redshift galaxies.
Combining a discrete galaxy population with approximate, efficient radiative transfer allows us to self-consistently forward-model galaxy surveys, line intensity maps (LIMs), and observations of the intergalactic medium (IGM). Not only does each observable probe different scales and physical processes, but cross-correlation will maximise the information gained from each measurement by probing the galaxy-IGM connection at high-redshift. We find that a stochastic source field produces significant shot-noise in 21cm and LIM power spectra. Scatter in galaxy properties can be constrained using UV luminosity functions and/or 21cm power spectra, especially if astrophysical scatter is higher than expected (as might be needed to explain recent {\it JWST} observations). Our modelling pipeline is both flexible and computationally efficient, facilitating high-dimensional, multi-tracer, field-level Bayesian inference of cosmology and astrophysics during the first billion years.}


\maketitle
%

\section{Introduction}
The Epoch of Reionisation (EoR) and Cosmic Dawn (CD) mark the formation of the first galaxies, whose multi-wavelength radiation heated and ionised the pervasive intergalactic medium (IGM). The history and morphology of this IGM phase transition encode a wealth of information about cosmology, large-scale structure formation and the astrophysics of the first galaxies.

Current and upcoming observations of the EoR and CD come from a wide variety of sources. The  brightest few galaxies are being seen individually by optical and infrared telescopes, such as the {\it Hubble} space telescope\footnote{\href{hubblesite.org}{hubblesite.org}}, the James Webb Space Telescope ({\it JWST})\footnote{\href{webbtelescope.org}{webbtelescope.org}}, the Very Large Telescope (VLT)\footnote{\href{eso.org/public/teles-instr/paranal-observatory/vlt}{eso.org/public/teles-instr/paranal-observatory/vlt}}, and in the future by the Nancy Grace Roman space telescope\footnote{\href{roman.gsfc.nasa.gov}{roman.gsfc.nasa.gov}} and the Extremely Large Telescope (ELT)\footnote{\href{elt.eso.org}{elt.eso.org}}. Line intensity mapping (LIM) experiments such as the ATacama Large Aperture Submilimetre Telescope ({\it AtLAST})\footnote{\href{atlast.uio.no}{atlast.uio.no}} and the Fred Young Sumbilimetre Telescope ({\it FYST})\footnote{\href{ccatobservatory.org}{ccatobservatory.org}}, and the Spectro-Photometer for the History of the Universe, Epoch of Reionization and Ices Explorer (SPHEREx)\footnote{\href{https://spherex.caltech.edu/}{https://spherex.caltech.edu/}} will observe the large-scale, integrated flux of {\it all} galaxies, using transitions in the interstellar medium (ISM) such as CO, CII and Lyman alpha (e.g. see the review in \citealt{Bernal22}). The Lyman alpha forest in the spectra of high-$z$ QSOs already gives strong constraints on IGM properties at $z<6$ (e.g. \citealt{Dodorico23}, \citealt{Qin24inference}).  Arguably the most revolutionary probe will come from the spin-flip transition of HI.  This 21cm line will allow us to eventually map the IGM during the first billion years, resulting in a cosmological dataset of unprecedented size and quality (e.g. see the review in \citealt{Mesinger19}).

These multi-tracer observations are highly complementary.   Combining them in a self-consistent analysis framework is fundamental for extracting the maximum cosmological and astrophysical information.  This is especially true with preliminary, low signal-to-noise (S/N) EoR/CD datasets, which {\it must} be combined in order to meaningfully constrain theoretical models (e.g. see Figure 8 and associated discussion in \citealt{Breitman24}).
Analysing diverse datasets will also allow us to identify possible tensions between observations, which can act as signposts for new physics and/or systematics.

Unfortunately, it is not easy to construct a self-consistent, flexible modeling framework for multi-tracer EoR/CD observations. Galaxy formation is a highly non-linear processes, with cosmic radiation fields coupling a vast range of scales.
Furthermore, the unknown physics of early galaxy formation means that models calibrated to low redshift data can give very different predictions at high redshifts and/or small masses where current data is lacking (e.g \citealt{Ni23}, \citealt{Lovell24}).
As a result, theoretical modeling pipelines of EoR/CD datasets need to be accurate, flexible, and fast enough to explore the large parameter space of uncertainties.

These requirements motivated development of so-called, "semi-numerical" simulations (e.g. \citealt{Mesinger07, Visbal12, Mutch16, Croton16, Choudhury18, Hutter21}). Semi-numerical simulations make trade-offs between accuracy and speed so that they can be used in data-driven analyses from a combination of IGM and galaxy observations (e.g. \citealt{Greig15,Choudhury21,Qin21,HERA22,Nikolic23,Mutch24}). Starting from large-scale, 3D realizations of initial conditions, linear/quasi-linear evolution is often performed using higher-order perturbation theory (e.g. \citealt{Scoccimarro98}) to avoid costly N-body simulations, while non-linear structure formation may be captured with an excursion-set approach (e.g. \citealt{Mesinger07}).  Galaxy properties are parametrized with a choice of physical or empirical functional forms.  The corresponding emissivities are used to perform multi-band cosmological radiative transfer using approximations that are reasonably accurate on moderate to large scales ($\gtrsim$ cMpc; e.g. \citealt{Mesinger11, Zahn11, Ghara18, Hutter18}).  As a result, semi-numerical simulations can generate 3D lightcones of various IGM properties in roughly 1 core hour: many orders of magnitude faster than more detailed cosmological radiative transfer simulations.

However, some approximations that are currently made in semi-numerical simulations could limit their usefulness when interpreting certain observations.  For example, a common assumption is that each $\sim$ cMpc simulation cell contains the {\it average} conditional halo mass function and corresponding galaxy emissivity, given its density, velocity, temperature, and ionization state.
This is typically justified by the fact that radiation fields and associated IGM properties are determined by the combined radiation of many sources.  In some cases however, neglecting galaxy-to-galaxy scatter could bias our interpretation of EoR/CD data (see e.g. \citealt{Ren19,Gelli24,Nikolic24}).

Furthermore, some observations might require modeling {\it individual} galaxies and their surrounding EoR morphology (e.g. \citet{Lu24,Nikolic25}).  Modelling individual galaxies in large-scale simulations currently requires either expensive N-body codes (e.g. \citealt{Choudhury18,Ghara23,Schaeffer23}) or somewhat faster Lagrangian halo finders (e.g. \citealt{Monaco02,Mesinger07}); these add significant computational and memory overheads limiting their use in Bayesian inference pipelines.

\begin{figure*}
    \centering
    \includegraphics[width=0.9\linewidth]{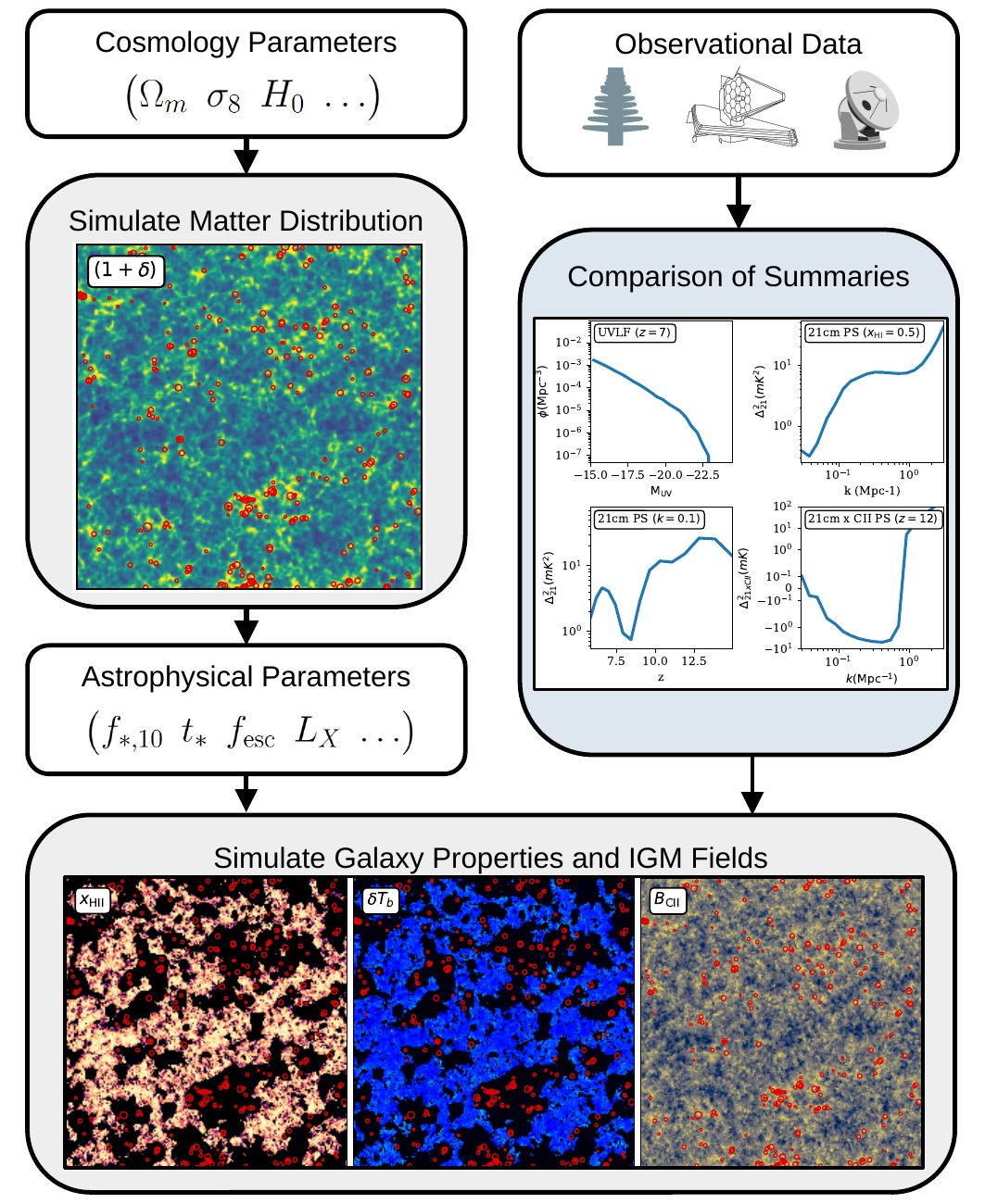}
    \caption{An example usage case for \cmfastvfour. First, cosmological parameters are sampled, and an associated realization of the Lagrangian matter field is generated.  Dark matter halos are identified in Lagrangian space and then moved, together with the matter field, to Eulerian space using 2LPT.  This is shown in the left-middle panel as a slice through a $300 \mathrm{Mpc}$ overdensity field at $z=7$, with the 300 largest halos in the 2 Mpc-thick slice shown as red circles. Then galaxies are assigned to DM halos by sampling parametric conditional probability densities based on well-established empirical relations such as the SHMR, SFMS, FMR, etc. Cosmic radiation fields (Lyman-alpha, Lyman-continuum, and X-ray) sourced by these galaxies are calculated using approximate radiative transfer, and the IGM is evolved accordingly.  The lower panel shows slices corresponding to the IGM neutral fraction, 21cm brightness temperature, and CII surface brightness density, at the same redshift as the overdensity field and with the same halos overlaid as red circles. We can then extract statistics of these fields for comparison with multi-tracer observations. In the centre-right panel we show the UV luminosity function at $z=7$, the 21cm power spectrum at the midpoint of reionization, the evolution of the 21cm power spectrum at $k=0.1 \mathrm{Mpc}^{-1}$, and the CII x 21cm cross-power spectrum at $z=12$.  Under fiducial settings, {\tt 21cmFASTv4} computes all of these steps in $\lesssim$3 core hours for a single realization, thus facilitating high-dimensional, multi-tracer, field-level Bayesian inference of cosmology and astrophysics during the EoR/CD.}
    \label{fig:fmschematic}
\end{figure*}

In this work, we introduce a fast, stochastic source model implemented in the new version of the public code \cmfastvfour\footnote{\href{github.com/21cmFAST/21cmFAST}{github.com/21cmFAST/21cmFAST}}.
Using a combination of Lagrangian halo finding and coarse time-step merger trees, we rapidly build 3D realizations of dark matter halos throughout the EoR/CD.  These are then populated with galaxies by sampling well-established empirical relations, such as the stellar-to-halo relation (SHMR), the star-forming main sequence (SFMS), and the fundamental mass metalicity relation (FMR).  The parameters of these relations (e.g. mean, scatter, temporal correlation) form a flexible, easy-to-interpret astrophysical basis, allowing us to {\it infer them from observations}.
\cmfastvfour\ is now able to explicitly account for stochastic galaxy formation, as well as generate 3D lightcones of galaxy properties (e.g. stellar mass, star formation rate) together with the corresponding IGM lightcones.
As illustrated in Fig. \ref{fig:fmschematic}, this allows us to rapidly and self-consistently forward-model galaxy maps, LIMs, IGM observations, as well as the corresponding cross-correlations.

This paper is organised as follows: section \ref{sec:halosampling} describes how we sample halo masses in a simulation volume. Section \ref{sec:galaxies} details how galaxy properties such as stellar mass, SFRs and metalicities are assigned to each halo. Section \ref{sec:radiation} describes the calculation of radiative backgrounds from the new discrete sources and their effect on the IGM. Section \ref{sec:example} presents results from a fiducial simulation, showing how the new source model drives the EoR/CD. Section \ref{sec:comparison} compares our new source model to previous iterations in \cmfast, detailing the effects the stochastic source model has on our lightcones. Finally, Section \ref{sec:forward} shows new outputs of the model which can be used for analysis, using the cross-correlation between the CII and 21cm intensity maps as an example. We then summarise our work and possible next steps in section \ref{sec:conclusion}. Throughout this paper we use comoving units unless stated otherwise, as well as the following cosmological parameters: $\left\{ \Omega_M, \Omega_b, H_0, \sigma_8 \right\} = \left\{ 0.3096 , 0.04897, 67.66, 0.8102 \right\}$.

\section{Generating the dark matter halo field}\label{sec:halosampling}
Our goal is to include discrete halos within an inference pipeline that can be sampled rapidly with different initial conditions.  N-body codes would be too computationally expensive, given the enormous dynamic range requirements: the first galaxies are hosted by halos with masses of
$\sim 10^6 M_\odot$, and the volumes required to sample the largest relevant scales during the EoR/CD are $\gtrsim (250$ Mpc)$^3$ \citep{Iliev14,Kaur20}.  

To achieve these requirements, various approximate approaches for populating cells with halos have been explored. The simplest is to integrate over a conditional halo mass function (CHMF) either within cells \citep{Davies22,Trac22,LIMFAST} or within larger spheres \citep{Mesinger11}. While these methods are fast, they ignore stochasticity in the halo mass distribution, which could be important for some observables (e.g. \citealt{Nikolic24}). If we wish to take into account the scatter in the halo mass distribution from the CHMF, we need to generate a population. The first methods for generating these populations rely on the excursion set formalism \citep{Bond91}, where a random walk is performed on the Lagrangian density field in decreasing smoothing scales, and halos form where this smoothed density crosses a redshift-dependent (and possibly scale dependent) barrier. Such methods require very high resolution density fields to resolve the smallest halos (e.g. \citealt{Monaco02, Mesinger07}), which can be very expensive in terms of memory and computational requirements.

Other methods follow binary mergers within a halo mass distribution using small internal time-steps \citep{Cole00,Parkinson08,Benson16,Qiu21,Trinca22}. Such binary merger algorithms can result in halo populations consistent with N-body simulations; however, they require "ad-hoc" calibration to match a target distribution
(see \citet{Cole08} and appendix \ref{app:sampler}). While it is possible to fit these calibration factors for any particular simulation or mass function, it would be desirable to have a more flexible model that can quickly produce samples from \textit{any} given CHMF without needing to perform a fit beforehand.

\begin{figure*}
	\includegraphics[width=\linewidth]{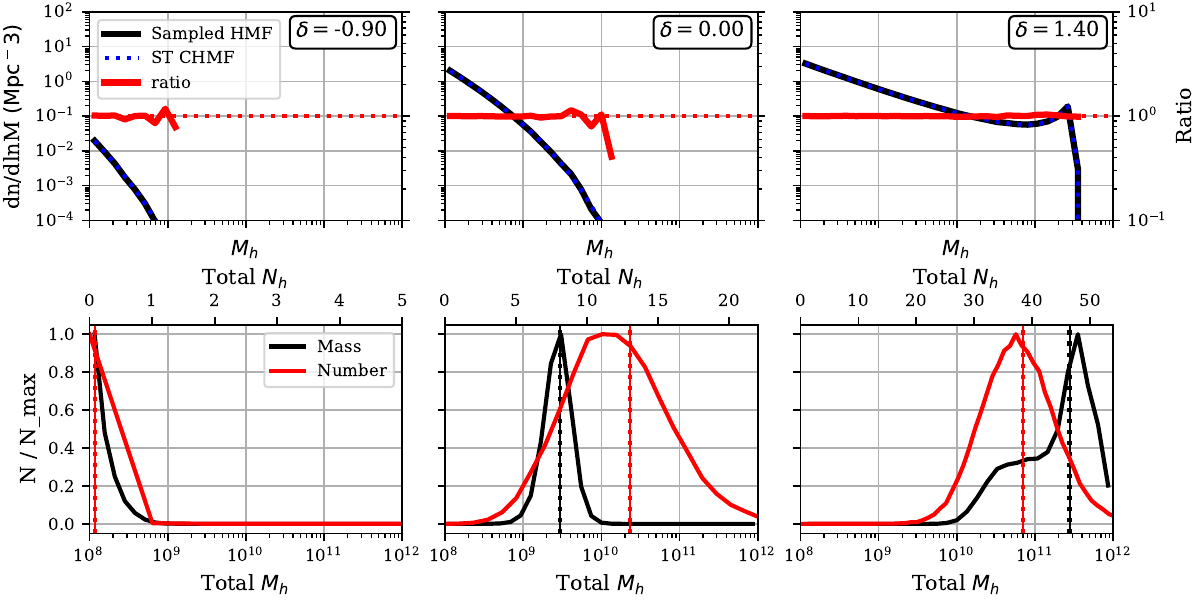}
    \caption{{\it Top row}: Conditional halo mass functions from our sampler at $z=10$ (solid black line) and the Sheth-Tormen CHMF (eq. \ref{eq:smtcmf}) (dotted blue line) in 10,000 underdense (left, $\delta = -0.9$) mean density (middle, $\delta = 0.0$) and overdense (right, $\delta = 1.4$) Lagrangian cells of volume (2Mpc)$^3$. Their ratio is indicated on the right vertical axis and shown as a solid red line which follows the target of unity ({\it dotted red}). {\it Bottom row}: Distribution of the number of halos (red curve) and total mass of halos (black curve) within each cell. The average of the distribution and the expected average from the Sheth-Tormen CHMF are shown as solid and dotted vertical lines, respectively. The agreement between the mean computed from samples and the target mean is evidenced by the overlapping solid and dotted lines.}
    \label{fig:firstsample}
\end{figure*}

Here we use a hybrid approach (c.f. \citealt{McQuinn07,Nasirudin20,Meriot24}) Halos whose mass are comparable or larger than the Lagrangian mass of a simulation cell are identified using the existing Lagrangian halo finder of {\tt 21cmFAST} titled \textsc{DexM} \citep{Mesinger07}.  Less massive halos are instead identified with new, coarse-step merger trees that sample from conditional halo mass functions (similar to \citealt{Sheth99,McQuinn07}).  We describe the different ingredients of this new method below.

\subsection{Conditional halo mass functions}
In order to distribute halos within a simulation volume more quickly, we use the conditional halo mass function $\frac{dn}{dM_h}(M_h, z | M_\mathrm{cond},\delta_\mathrm{cond})$, which denotes the number density per unit mass of halos of mass $M_h$ at redshift $z$, given that they reside within a Lagrangian volume of mass $M_\mathrm{cond}$ and mean overdensity $\delta_\mathrm{cond}$. The only fully analytic CHMF in widespread use is the Extended Press-Schechter (EPS) conditional mass function \citep{Lacey93} which computes the halo mass function from the first time a random walk in decreasing Lagrangian scale, $M_h$, crosses a critical value of the matter overdensity, $\delta_\mathrm{crit}= 1.686/D(z)$, where $D(z)$ is the linear growth factor.   For CHMFs, the origin of the random walk is the value $\delta_\mathrm{cond}$ at scale $M_\mathrm{cond}$, and the root-mean-square fluctuation of the linear matter field at $z=0$, $\sigma(M_h)$, is set by the assumed cosmology.

The halo mass distribution resulting from this model is
\begin{multline}\label{eq:epscmf}
    \frac{dn_\mathrm{EPS}}{dM_h}(M_h,z|\sigma_\mathrm{cond},\delta_\mathrm{cond}) = \frac{\rho_{\mathrm{crit}} \Omega_M}{\sqrt{2\pi} M_h} \frac{2\sigma (\delta_\mathrm{crit} - \delta_\mathrm{cond})}{(\sigma^2 - \sigma_\mathrm{cond}^2)^{3/2}} \\ 
    \left| \frac{d\sigma}{dM_h} \right|  \mathrm{exp} \left[ -\frac{(\delta_\mathrm{crit} - \delta_\mathrm{cond})^2}{2(\sigma^2 - \sigma_\mathrm{cond}^2)} \right]
\end{multline}
where $\rho_{\mathrm{crit}}$ is the critical matter density and we denote $\sigma(M_h) \rightarrow \sigma$ and $\sigma(M_\mathrm{cond}) \rightarrow \sigma_\mathrm{cond}$ for brevity. While this model can describe the hierarchical formation of halos over cosmic time, it under-predicts the number of massive halos and over-predicts the number of small mass halos, when compared to N-body simulations (e.g \citealt{Tinker08}). As a result, various methods are applied to correct this function, including altering the shape of the barrier $\delta_\mathrm{crit}$ \citep{Sheth01}, scaling the CHMF to match total collapsed fraction estimates \citep{Barkana04,Mesinger11}, adding corrective terms which scale with $\delta$ and $\sigma$ \citep{Parkinson08}, or scaling an unconditional mass function by performing variable substitutions (similar to the $\delta_\mathrm{crit} \rightarrow (\delta_\mathrm{crit} - \delta_\mathrm{cond})$ used to compute the EPS CHMF from the unconditional one; \citealt{Rubino08,Tramonte17,Trapp20}.

\begin{figure*}
	\includegraphics[width=\linewidth]{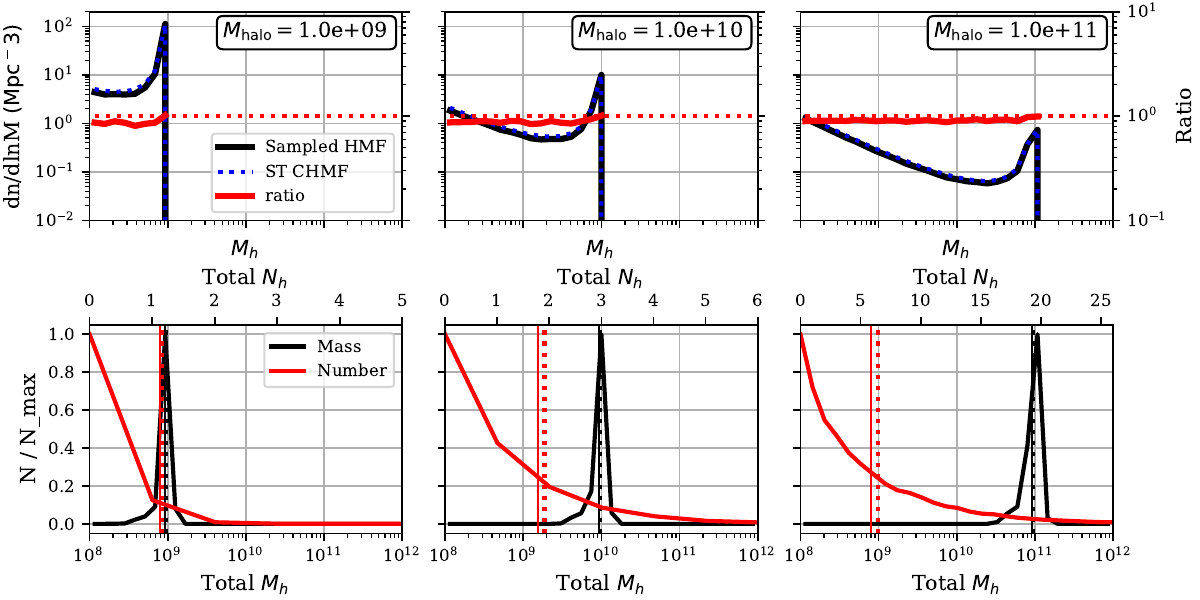}
    \caption{Same as Figure \ref{fig:firstsample}, but sampling from 10,000 descendant halos of mass $10^9$, $10^{10}$ and $10^{11} M_\odot$ from redshift $z_\mathrm{desc}=6$ to the progenitor redshift $z_\mathrm{prog} = (1+z_\mathrm{desc})\Delta_z - 1$ where $\Delta_z = 1.02$. Samples drawn from descendant halos use the mass-limited sampling described above, rather than the number-limited sampling used for the cells. The short time-step results in a very tight distribution of total progenitor mass, although the number of halos and their mass distribution can vary significantly between individual descendants.}
    \label{fig:progsample}
\end{figure*}

Here we use the conditional mass function from \citet{Sheth02}, which is approximated using a Taylor expansion of the critical density threshold associated with ellipsoidal collapse
\begin{equation}
    B(z,\sigma) = \sqrt{a} \delta_\mathrm{crit} \left( 1 + \beta \left( \frac{a\delta_\mathrm{crit}^2}{\sigma^2} \right)^{-\alpha} \right)
\end{equation}
where we use the parameters from \citet{Jenkins01} $a = 0.7$, $\alpha = 0.81$, $\beta = 0.34$. The conditional mass function is then well approximated by
\begin{multline}\label{eq:smtcmf}
    \frac{dn_\mathrm{ST}}{dM_h}(M_h,z|\sigma_\mathrm{cond},\delta_\mathrm{cond}) = \frac{\rho_{\mathrm{crit}} \Omega_M}{\sqrt{2\pi}M_h} |T(z,\sigma,\sigma_\mathrm{cond})| \\
    \frac{d\sigma}{dM_h} \frac{2\sigma}{(\sigma^2 - \sigma_\mathrm{cond}^2)^{3/2}}
    \mathrm{exp} \left[ \frac{-(B(z,\sigma) - \delta_\mathrm{cond})^2}{2(\sigma^2 - \sigma_\mathrm{cond}^2)} \right]
\end{multline}
where
\begin{equation}
    T(z,\sigma, \sigma_\mathrm{cond}) = \sum_{n=0}^5 \frac{-(\sigma^2-\sigma_\mathrm{cond}^2)^n}{n!} \frac{\partial^n}{\partial(\sigma^2)^n} \left( B(z,\sigma) - \delta_\mathrm{crit} \right)
\end{equation}

While we assume the Sheth-Tormen CHMF for the results shown in this paper, the sampling method implemented in \cmfastvfour\ and detailed in the remainder of this section is applicable to any user-defined CHMF.

\subsection{Sampling the CHMFs}\label{sec:sampling}
We build on the \cmfast\ framework by including a fast stochastic source sampler capable remaining self-consistent across cosmic time.
We begin our sampling at the lowest redshift requested by the user, where each cell in the linearly evolved Lagrangian density grid is used as a condition, determining the CHMF from which we sample using equation \ref{eq:smtcmf}. $M_\mathrm{cond}$ corresponds to the Lagrangian mass of the cell, and $\delta_\mathrm{cond}$ is set to the density of our initial conditions field linearly evolved to $z=0$.
For computational efficiency, we build an interpolation table of the inverse cumulative distribution function of the CHMF and sample it in the mass range $[M_\mathrm{min},M_\mathrm{cell}]$, where $M_\mathrm{min}$ is a free parameter describing our halo mass resolution. We sample the CHMF of each cell $N_h$ times, where $N_h$ itself is drawn from a Poisson distribution with mean
\begin{equation}\label{eq:numlim}
\bar{N}_h = V_\mathrm{cell}\int_{M_\mathrm{min}}^{M_\mathrm{cell}} \frac{dn}{dM_h}(M_h,z ~|~\sigma_\mathrm{cond},\delta_\mathrm{cond}) dM_h
\end{equation}

In Figure \ref{fig:firstsample} we demonstrate the accuracy of this method on the grid scale by showing the mass functions drawn for 10,000 cells of side length $2$ Mpc and differing Lagrangian overdensity, compared to the expected conditional mass functions. Both the total number and total mass of halos above the minimum mass can vary significantly between samples.

Since the mass of the cell is determined entirely by the cell size which is uniform throughout the simulation, it is impossible to sample halos with a higher mass from its CHMF. We use a cell length of $2$ Mpc throughout this work, which includes halos up to a mass of $3.11 \times 10^{11} M_\odot$.
To find halos above this mass, we utilise the existing halo finder {\tt DexM} \citep{Mesinger07}.  {\tt DexM} works by filtering the initial condition density grid on a series of decreasing scales, identifying halos by the scale at which the Lagrangian density crosses a certain barrier and accounting for mass conservation. In order not to double-count halos that are part of larger structures identified by {\tt DexM},  we multiply $\bar{N}_h$ in each cell by one minus the fraction of mass taken up by these larger halos before sampling the CHMF. Using {\tt DexM} only once per simulation (as opposed to every redshift) and only to find the largest halos allows us to sample the full halo mass range within our box while retaining the computational efficiency gained by using the sampler on smaller scales.

If we were to simply repeat this process at each redshift, each halo population would follow the CHMF of the Lagrangian cell in which it resides and would contain the correct mean number and total mass of halos. However, samples in adjacent snapshots would not be correlated with one another, and as a result the halo population in a given region would fluctuate wildly over short periods of time. Since we aim to build lightcones of our halo populations and the resulting radiation fields, we need to correlate the halos sampled at each redshift. To do this, we take our first sample from the lowest redshift as described above, then step back from the descendant redshift $z_\mathrm{desc}$ to a progenitor redshift $z_\mathrm{prog}$ (where $z_\mathrm{desc} < z_\mathrm{prog}$).  We treat each halo as an independent Lagrangian volume and sample from its CHMF setting the condition mass to the descendant halo mass as $M_\mathrm{cond}=M_\mathrm{desc}$ and setting the condition density to the SMT barrier at the descendant redshift $\delta_\mathrm{cond} = B \left(z_\mathrm{desc},\sigma_\mathrm{cond} \right)$\footnote{This value is specific to the SMT mass function. Generally speaking, it is the condition for halo formation at the descendant redshift (possibly dependent on halo mass, redshift or other factors), evolved to the progenitor redshift.} in equation \ref{eq:smtcmf}. In the excursion set framework this is analogous to starting the random walk where the descendant halo crosses the barrier $B \left( z_\mathrm{desc}, \sigma_\mathrm{cond} \right)$ at the mass scale $M_\mathrm{desc}$ and sampling the mass $M$ where paths from this point first cross the higher barrier $B \left( z_\mathrm{prog}, \sigma \right)$.

Due to the short time-step and smaller condition masses, the Poisson distribution is no longer a good approximation for the number of halos obtained from a descendant, which tends to be much more narrowly distributed around unity. The use of the Poisson distribution to find progenitors would cause halos to disappear entirely or double in mass on very short timescales. Therefore instead of limiting the sample by number, we limit it by mass according to the expected mass of progenitors above a limit $M_\mathrm{lim}$ set by the conditional mass function.
\begin{equation}\label{eq:masslim}
M_\mathrm{lim} = \frac{M_\mathrm{cond}}{\rho_\mathrm{crit}\Omega_m}\int_{M_\mathrm{min}}^{M_\mathrm{cond}} M_h \frac{dn}{dM_h}(M_h,z~|~\sigma_\mathrm{cond},\delta_\mathrm{cond}) dM_h
\end{equation}
This effectively assumes that in each descendant halo, a mass of $M_\mathrm{cond} - M_\mathrm{lim}$ resides in objects below our mass resolution (i.e. obtained by smooth accretion), which are ignored in further sampling.

Once this mass limit is exceeded, the final halo is included if it brings the total mass closer to our target mass $M_{\mathrm{lim}}$, allowing for some variance in the total progenitor mass while favouring samples which are closest to the total mass expected\footnote{While it may seem reasonable to enforce a more strict level of mass conservation within each descendant, in practice this skews the resulting mass functions, as certain samples are more likely to fall within a given mass tolerance. Progenitor samples are not directly correlated with one another, so we would see a large deficit of halos around $M_\mathrm{prog} = M_\mathrm{desc}/2$ since samples including halos at this mass are far less likely to meet any mass tolerance criteria.}. This process is repeated in successive backward time-steps of $z_\mathrm{prog} = (1+z_\mathrm{desc})\Delta_z - 1$, where $\Delta_z$ is a free parameter set to $1.02$ for results shown in this paper. We show resulting distributions from our mass-limited sampling in Figure \ref{fig:progsample}, where we observe a slight deficit (around 5\%) of smaller progenitors, meaning that although the total mass of progenitors is correct, our model will produce slightly fewer merger events than the CHMF predicts. Our mass-limited sampling is very similar to the mass partitioning method presented in \citep{Sheth99}, but instead of sampling from the cumulative collapsed mass function and progressively updating the condition mass and collapse barrier, we draw all samples from the same CHMF at $M_\mathrm{desc}$ and $B \left( \sigma(M_\mathrm{desc}) \right)$. These modifications were made for easy tabulation and fast sampling of a general CHMF.

We provide comparisons between these methods, as well as further details regarding our model choice in Appendix \ref{app:sampler}. The mass partitioning algorithm from \citet{Sheth99} as well as the binary-split merger tree algorithm from \citet{Parkinson08,Qiu21} are also available as sampler options within \cmfastvfour.

\begin{figure}
	\includegraphics[width=\linewidth]{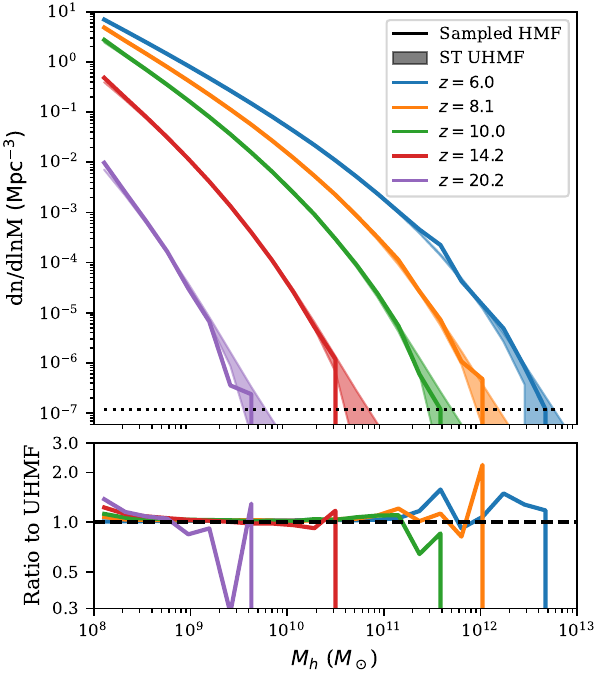}
    \caption{The correlated sampling of halos from $z=6$ to $z=30$ within a $300$ cMpc box. Top panel: The total halo mass function within the box at from redshifts $z=6$ to $z=30$, compared to the Sheth-Tormen unconditional mass function represented as a shaded region with Poisson uncertainty. The black dotted line corresponds to a number density of one halo in the simulation volume within that bin. Bottom Panel: The ratio of the sampled mass function to the expected Sheth-Tormen unconditional mass function, showing an excess of small halos at the highest redshifts, but remaining within Poisson errors over the vast majority of masses and redshifts.}
    \label{fig:boxupdate}
\end{figure}

We show summary results of our halo sampling algorithm in Figure \ref{fig:boxupdate}, which contains the halo mass functions in a 300 Mpc box produced by our hybrid approach of DexM Lagrangian halo finding combined with the new coarse time step merger tree algorithm.  The number-limited sampling to identify the descendents for the merger trees was performed at $z=6$, followed by mass-limited sampling of progenitors up to $z=30$ in steps of $z_\mathrm{prog} = (1+z_\mathrm{desc})\Delta_z - 1$ where $\Delta_z = 1.02$. The bottom panel shows the ratio of the halo mass functions to the target unconditional Sheth-Tormen mass function.  At the smallest scales we see a slight excess of halos produced by the sampler, which grows to  $20\%$ at the highest redshifts $z \gtrsim 20$; however, this does not greatly affect the radiation fields as the source population is very small at these times.  The shaded regions in the top panel show the $1\sigma$ Poisson noise.  Overall, our halo mass functions are within the Poisson noise over six orders of magnitude down to the single halo limit.

Halos in the initial number-limited sample are placed at a random location within their Lagrangian cells, which are mapped onto the higher-resolution grid of the initial conditions. Progenitors of each halo from the mass-limited sampling are then placed within the same cell on this higher-resolution grid as their descendants. In order to place each halo in its correct location at a particular redshift, second-order Lagrangian Perturbation Theory is used \citep{Scoccimarro98}. As a result, halos are moved alongside the underlying density field, and are allowed to cross cell boundaries.
While our model is orders of magnitude less computationally expensive than comparable N-body simulations with halo finders, it does add non-trivial computational time and memory usage to \cmfast, this can be roughly $2 - 3$ times the run time, dependent on parameter settings.

It is important to note however that for most usage cases, not all halos need to be resolved.  We account for the radiation emitted from halos below the user-defined limit by integrating over the average CHMF between the sampler minimum mass and a cooling scale below which no stars form.  This neglects halo stochasticity below the sampler mass limit, though we find $\lesssim 10$ percent level differences in the resulting IGM properties when computing emissivities using average number densities below $\lesssim10^{9} M_\odot$ compared to resolving all of the halos all the way down to the atomic cooling threshold.  Given that galaxies inside $\lesssim10^{10} M_\odot$ halos are generally too faint to be detected individually, even with {\it JWST} (e.g. \citealt{Harikane23}), a user wishing to forward-model galaxy and corresponding IGM fields could choose a higher value for the sampler minimum mass and still obtain accurate realizations of both observable galaxies and IGM properties.  Such a set-up (c.f. Triantafyllou, in prep) would reduce the memory/computational requirements to levels similar to \cmfast v3.
Further details on performance and memory usage can be found in Appendix \ref{app:time}.

\section{Populating halos with galaxies}\label{sec:galaxies}
Assigning galaxies to host dark matter halos is a difficult task, especially at high redshifts where observational data are relatively sparse and the physics of star formation, feedback and galaxy evolution is even more uncertain.  Current approaches typically rely either on hydrodynamic simulations, semi-analytic or semi-empirical models.  Each have their own benefits and drawbacks.

Hydrodynamic simulations are appealing due to the fact that they capture gas dynamics and the gravitational interplay of baryons and dark matter.  However, cosmological simulations cannot resolve individual star formation events and the associated feedback, and therefore must rely on resolution-dependent, sub-grid prescriptions that are calibrated against observations.  Unfortunately the predictions of such sub-grid recipes can vary dramatically in the regime where there is currently no data (e.g. \citealt{Ni23,Lovell24}).  They are also numerically expensive and so cannot be directly used in an inference framework.

Semi-analytic models (SAMs) on the other hand connect galaxies to halos using analytic equations for gas cooling, star formation and feedback (e.g. \citealt{Croton16,Mutch16,Hutter21}).  They are reasonably fast to evaluate.  However, they typically include several "baked-in" assumptions that might not hold for high redshift galaxies, such as spherical symmetry, rotation-supported disks, etc.  Moreover, the free parameters of a given SAM can be difficult to interpret and generalize to other models.

Semi-empirical models use parametric relations to connect galaxies to halos, fitting these relations to available data (e.g. \citealt{Vale06,Behroozi19,Park19}).  This makes them fast and generalizable, at the cost of losing direct insight into the physics regulating star formation and galaxy evolution.  Moreover, there is no unique choice of which relations to parametrize and with which functional forms.

As {\tt 21cmFASTv4} is intended to be used in a forward-modelling framework, we have several requirements for our halo--galaxy connection:
\begin{packed_enum}
\item {\it Speed} -- it must be fast enough for Bayesian inference.
\item {\it Transparency}  -- the galaxy--halo connection should be explicitly-defined, facilitating transparency of the model and its assumptions.
\item {\it Flexibility} -- given our relatively poor knowledge of star formation and galaxy evolution in the first billion years, our model should be flexible enough to capture a large range of scenarios.
\item {\it Generalizability} -- the galaxy--halo connection should be made in a "universal language" that can accommodate diverse theoretical models as well as observations, allowing us to set well-motivated priors on its free parameters.
\end{packed_enum}

\begin{table}
    \centering
    \begin{tabular}{ c|c|c }
        \hline
         Parameter & Description & Value \\
         \hline
         $f_{*,10}$ & SHMR normalisation & 0.05 \\
         $\alpha_*$ & Low-mass scaling of SHMR & 0.5 \\
         $\alpha_{*2}$ & High-mass scaling of SHMR & -0.61 \\
         $M_\mathrm{pivot}$ & Pivot mass in SHMR & $2.8 \times 10^{11} M_\odot$ \\
         $\sigma_*$ & Scatter in SHMR & 0.3 dex \\
         $C_*$ & Correlation time of SHMR & 0.5 \\
         $t_*$ & SFMS normalisation & 0.5 \\
         $\sigma_\mathrm{SFR,lim}$ & Scatter limit in SFMS & 0.19 dex \\
         $\sigma_\mathrm{SFR,idx}$ & $M_*$ scaling of SFMS scatter & -0.12 \\
         $C_\mathrm{SFR}$ & Correlation time of SFMS & 0.1 \\
         $E_0$ & X-ray energy threshold & 0.5 keV \\
         $\alpha_X$ & X-ray spectral index & 1 \\
         $L_{X,\mathrm{norm}}$ & $L_X / \mathrm{SFR}$ normalisation & $10^{40.5} \mathrm{erg s}^{-1}$ \\
         $\sigma_X$ & scatter in $L_X / \mathrm{SFR}$ & 0.5 dex \\
         $C_\mathrm{X}$ & Correlation time of $L_X / \mathrm{SFR}$ & 0.5 \\
         $f_\mathrm{esc,10}$ & $f_\mathrm{esc}$ normalisation & 0.1 \\
         $\alpha_\mathrm{esc}$ & $f_\mathrm{esc}$ mass scaling & 0.5 \\
         $N_\gamma$ & LyC photons per stellar baryon & 5000
    \end{tabular}
    \caption{Fiducial parameters for galaxy scaling relations used in equations \ref{eq:lognorm} to \ref{eq:fesc}}
    \label{tab:galparams}
\end{table}

In order to accommodate these requirements, here we adopt a semi-empirical approach, but one that is anchored to well-studied galaxy scaling relations.  Specifically, we define {\it conditional probability densities} that stochastically connect galaxy properties to host halo masses.  These include the stellar-to-halo mass relation [SHMR; $P(M_*~|~M_h)]$, the star forming main sequence [SFMS; $P(\mathrm{SFR}~|~M_*, z)$], fundamental mass metallicity relation [FMZ; $P(Z ~|~ M_*,\mathrm{SFR})$], and the X-ray luminosity to star formation rate [LxSFR; $P(L_X ~|~ \mathrm{SFR}, Z)]$ \footnote{To be more precise, relations like the SHMR and the SFMS typically refer to the running averages of these conditional distributions, obtained by fitting a line in log-log space.  Here we sample from the full conditional probability densities, which include also scatter around the mean relations.}.   Sampling from conditional probability distributions is extremely fast, easy to interpret, flexible, and allows us to characterize a wide range of theoretical models and observational data.
Ultimately, we want to {\it infer the parameters of these distributions} (e.g.  means, scatters, temporal correlations, etc.) from multi-tracer observations\footnote{A multi-tracer approach is especially important during the EoR/CD, as existing data-sets are very complementary: direct galaxy observations with {\it Hubble/JWST} provide information about rare, bright galaxies while IGM and LIM observations are sensitive to the cumulative photon emission that is dominated by the abundant, faint galaxies far below detection limits of {\it JWST} (e.g. \citealt{Breitman24,Qin21minihalo}).}, using prior ranges informed by both simulations and low-redshift observations.  These inferred properties of the unseen first galaxies can subsequently be interpreted using dedicated galaxy simulations/models, in an analogous manner to how we interpret direct observations of local galaxies.  We detail our approach below.

After halo masses are calculated for a snapshot, we assign galaxy properties to each halo by sampling from conditional distributions with free parameters governing their means and scatters. Our distributions for each property are taken mostly from \citet{Nikolic24}, with small differences which are detailed below. The stellar mass, star formation rate, and X-ray luminosity of each halo are sampled from separate log-normal distributions
\begin{equation}\label{eq:lognorm}
    P(\log(x)) = \mathcal{N} (\mu_{\log x},\sigma_{\log x}) = \frac{1}
    {\sigma\sqrt{2\pi}}\mathrm{exp} \left[ -\frac{1}{2}\left(
    \frac{\log(x) - \mu_{\log x})}{\sigma_{\log x}} 
    \right)^2 \right]
\end{equation}
where $x$ is the halo property of interest, $\mu_{\log x}$ and $\sigma_{\log x}$ are the mean and standard deviation of the \textit{logarithm} of the property. It is noteworthy that the log-normal distribution is asymmetric, and the mean value of a variable drawn from the distribution is given by $\mu_{x} = 10^{\mu_{\log x} + \sigma_{\log x}^2/2}$ (see Appendix B of \citealt{Nikolic24}). In order to avoid a parametrisation scheme where changing the amount of scatter drastically alters the mean properties, by default we define our scaling relations in terms of their means $\mu_{x}$, rather than their mean logarithms $\mu_{\log x}$ as is the standard for observational fits. However, we investigate the latter models, which do not have a normalised mean, in section \ref{sec:comparison}.

We assign a correlation $\rho_{\log x}$ between progenitor and descendant properties dependent on the redshift step via a free parameter $C_x$:
\begin{equation}\label{eq:pcorr}
    \rho_{\log x} = \mathrm{exp} \left( -\frac{z_\mathrm{prog} - z_\mathrm{desc}}{C_x} \right)
\end{equation}
In order to generate samples with the target correlation, mean, and variance, we sample each progenitor property $\log(x_p)$ from the conditional bivariate Gaussian distribution given the descendant property $\log(x_d)$\footnote{As discussed in the previous section, our algorithm steps backwards in time and therefore the progenitor properties are sampled after those of the descendant.}, which itself is a Gaussian:
\begin{multline}
    P(\log (x_{p})) = \mathcal{N} \left( \mu_{\log x,p} + \rho_{\log x} \frac{\sigma_{\log x,p}}{\sigma_{\log x,d}}\left( \log(x_d) -\mu_{\log x,d} \right), \right. \\
    \left. \sqrt{1-\rho^2} \sigma_{\log x,p} \right)
\end{multline}
where subscripts $p$ and $d$ mark means and variances calculated using progenitor and descendant properties respectively. In the limiting case of $\rho_{\log x} = 0$, $\log(x_p)$ is simply sampled from $\mathcal{N}(\mu_{\log x,p},\sigma_{\log x,p})$ uncorrelated from its descendant. In the case of $\rho_{\log x}=1$, it is mapped directly from $\log(x_d)$, adjusting for the differences in mean and variance between the two distributions.

\begin{figure*}
	\includegraphics[width=\linewidth]{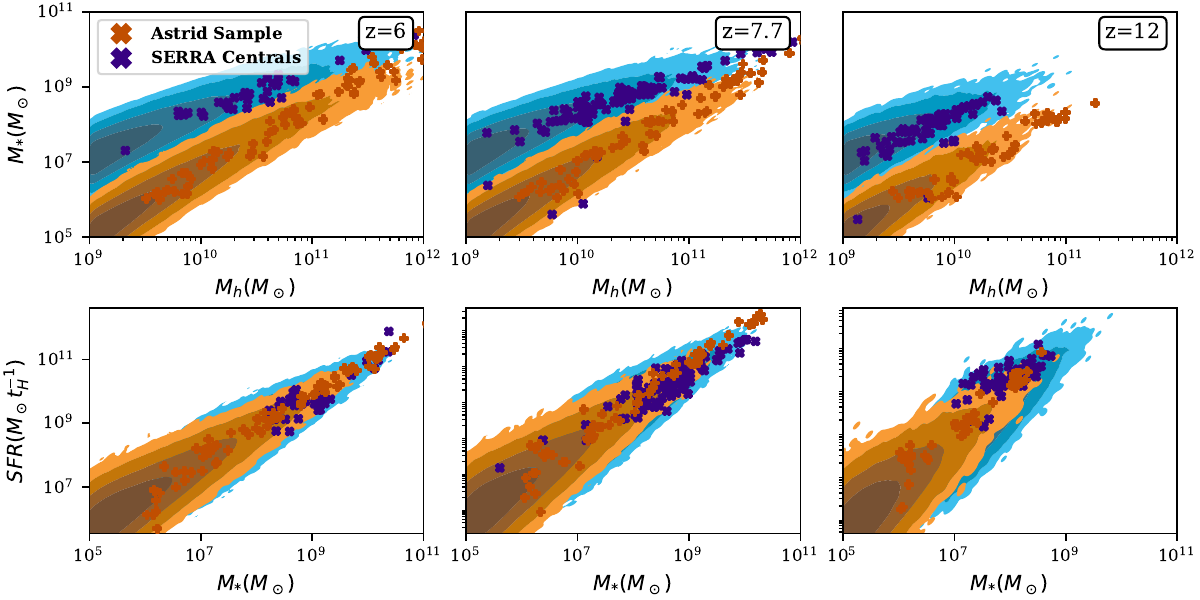}
    \caption{Distributions of stellar to halo mass ({\it top panels}) and star formation rate to stellar mass ({\it bottom panels}) at $z = (6.0,7.7,12)$. Orange and blue points correspond to galaxies taken from two hydrodynamical simulations: the zoom-in simulation suite SERRA \citep{Pallottini22} and the large cosmological simulation Astrid \citep{Bird22} For the SERRA sample we plot every central galaxy at these redshifts, while for Astrid we plot a subsample randomly selected in fixed logarithmic mass bins.
    Both codes have been calibrated to reproduce observable data at $M_h \gtrsim 10^{12} M_\odot$, but predict very different SHMRs for the unseen, faint galaxies that dominate the EoR and CD.
    The orange and blue contours correspond to two different parameter combinations in {\tt 21cmFASTv4} whose conditional distributions can characterize the correspondingly-coloured galaxy populations from the different hydro codes.  These contours correspond to 2--5 $\sigma$ of the {\it joint} distributions, $P(M_*, M_h)$ and $P({\rm SFR}, M_*)$, highlighting how the vast majority of galaxies are expected to be far below the resolution limits for large-scale cosmological simulations.
    This figure demonstrates that our semi-empirical model is flexible, capable of capturing a large range of predictions, and eventually allowing us to {\it infer} such galaxy properties from multi-tracer galaxy and IGM observations.
    }
    \label{fig:scaling}
\end{figure*}

We parametrize the mean and scatter of each conditional probability distribution using physically or empirically motivated functional forms.
The mean stellar mass at fixed halo mass $\mu_*$ is set by the following scaling relation \citep{Mirocha17}:
\begin{multline}\label{eq:acgstars}
    \mu_* = f_{*,10} \left(
    \frac{(M_\mathrm{pivot}/10^{10}M_\odot)^{\alpha_{*}} + (M_\mathrm{pivot}/10^{10}M_\odot)^{\alpha_{*2}}}{(M_h/M_\mathrm{pivot})^{-\alpha_{*}} + (M_h/M_\mathrm{pivot})^{-\alpha_{*2}}} \right) \\
    M_h
    \exp \left(\frac{-M_\mathrm{turn}}{M_h} \right) ~ 
    \frac{\Omega_b}{\Omega_m}
\end{multline}
where $f_{*,10}$ is the value of the SHMR at $10^{10}$ solar masses, and $\alpha_*$ is its low-mass power-law slope. We include an additional high-mass power-law component compared to \citet{Park19}, where the SHMR transitions to a power-law slope of $\alpha_{*2}$ at a pivot mass of $M_\mathrm{pivot}$. Previously, in {\tt21cmFAST} it was not important to include a separate high-mass slope in the SHMR since halo masses above $M_h \gtrsim 10^{11} M_\odot$ are too rare to contribute significantly to reionisation in data-constrained models (e.g. \citealt{Qin24inference}). Here however we include the turnover since {\tt 21cmFASTv4} is intended to forward-model galaxy surveys that can only observe such rare, bright galaxies \citep{GH25}. The characteristic halo mass below which star formation is exponentially suppressed, $M_\mathrm{turn}$, can either be used as a free parameter or it can be modelled as the largest of three physical scales  (c.f. \citealt{Qin20}): (i) $M_{\rm SNe}$, a free parameter corresponding to disruption from supernovae feedback; (ii) $M_{\rm cool}$,  a redshift-dependent atomic-cooling threshold corresponding to a virial temperature of $10^4$ K; and (iii) $M_{\rm photo}$, a characteristic scale set by photo-heating feedback that depends on the local reionization redshift and ionizing background of each cell \citep{Sobacchi13}. The standard deviation in the log stellar-to-halo mass ratio $\sigma_{*}$ is a free parameter, set in our fiducial model to a constant 0.3 dex, roughly motivated by the results of hydrodynamic simulations (e.g. \citealt{Hassan22,Nikolic24}).

The mean star formation rate (SFR) is determined by the following relation from \citet{Park19}:
\begin{equation}\label{eq:sfrmean}
    \mu_{\mathrm{SFR}} = \frac{M_*}{t_*H(z)} ~,
\end{equation}
where $M_*$ is the stellar mass of the galaxy (sampled from $P(M_* | M_h)$ as discussed above) and $t_*$ is a free parameter corresponding to the characteristic star formation time-scale in units of the Hubble time $1/H(z)$ (which also scales as the halo dynamical time during matter domination).  We allow for a parametric scatter around the mean SFMS, which by default decreases with increasing stellar mass according to a power-law:
\begin{equation}\label{eq:sigmasfr}
    \sigma_\mathrm{SFR} = \mathrm{max} \left( \sigma_\mathrm{SFR,lim},\sigma_\mathrm{SFR,idx}  \log \left( \frac{M_*}{10^{10}M_\odot} \right) + \sigma_\mathrm{SFR,lim} \right) ~.
\end{equation}
Here $\sigma_\mathrm{SFR,lim}$ is the high-mass limit of the SFMS scatter, and $\sigma_\mathrm{SFR,idx}$ is its power-law index with stellar mass.

Finally, the mean rest-frame soft-band (with photon energies less than 2keV) X-ray luminosity per unit star formation, $L_{\rm X, <2keV}$,  of high redshift galaxies is modelled as a double power-law dependent on the SFR and stellar mass, via the gas-phase metallicity $Z$:
\begin{equation}\label{eq:xraymean}
    \mu_{\mathrm{X}} = \frac{\mathrm{SFR}}{M_\odot \mathrm{yr}^{-1}} \times L_{X,\mathrm{norm}} \left( \left( \frac{Z}{0.05 Z_\odot} \right)^{0.64} + 1 \right)^{-1}
\end{equation}
where $Z$ is calculated using the FMR presented in \citet{Curti20}, and subsequently adjusted for redshift evolution  \citep{Curti24}:
\begin{equation}
    \frac{Z}{Z_\odot} = 1.23 \left( 1 + \left( \frac{M_*}{M_0} \right)^{-2.1} \right)^{-0.148}
    10^{-0.056z + 0.064}
\end{equation}
where
\begin{equation}
    M_0 = 10^{10.11} \left( \frac{\mathrm{SFR}}{M_\odot yr^{-1}} \right) ^{0.56}
\end{equation}
We use a double power-law to be consistent with high-metallicity measurements from \citet{Brorby16} with a flattening of $L_{\rm X, <2keV}/\mathrm{SFR}$ at low metallicities (e.g: \citealt{Fragos13,Lehmer21,Kaur22,Geda24}). We assume the specific X-ray luminosity is a power-law with photon energy $L_X \propto E^{-\alpha_X}$ where $\alpha_X$ is the spectral index of X-ray sources. The luminosity is normalised such that its integral is equal to the soft-band luminosity sampled above
\begin{equation}\label{eq:xrayintegral}
    L_{\rm X, <2keV}(M_*,\mathrm{SFR}) = \int_{E_0}^{2\mathrm{keV}}L_X(M_*,\mathrm{SFR}) dE
\end{equation}
Where $E_0$ is the energy threshold above which X-rays can escape their host galaxies.  Here use a fiducial value of $E_0 = 0.5$ keV, motivated by hydrodynamic simulations of the first galaxies \citep{Das17}.

While the temporal auto-correlation lengths, $C_x$, are free parameters within \cmfastvfour, their main purpose is to ensure that the variation in halo properties does not depend on the length of the simulation time-step. Default values for $C_*$ and $C_\mathrm{SFR}$ were set to be roughly consistent with simulated galaxies in Astrid \citep{Bird22}. We leave the exploration of these parameters to future work.

In contrast to \citet{Nikolic24}, we do not include scatter around the FMR, since this scatter was found to be sub-dominant in determining all radiation fields. At this stage, we also do not include scatter in the ionising escape fraction, as there is currently no consensus on how to parametrise it (e.g. \citealt{Yeh23,Kreilgaard24}).  Furthermore, scatter in the escape fraction is unlikely to impact large-scale radiation fields unless there is a strong dependence on galaxy properties (e.g. \citealt{Hassan22,Nikolic24}). Instead, we use the \citet{Park19} relation for the mean escape fraction which scales with halo mass:
\begin{equation}\label{eq:fesc}
f_\mathrm{esc} = f_\mathrm{esc,10} \left( \frac{M_h}{10^{10}M_\odot} \right)^{\alpha_\mathrm{esc}}
\end{equation}
which allows us to compute the cumulative number of ionizing photons which escaped from a given halo:
\begin{equation}\label{eq:nion}
    n_\mathrm{ion} = N_\gamma f_\mathrm{esc} M_*
\end{equation}
where we take the number of ionising photons per stellar baryon $N_\gamma = 5000$ for use in the excursion-set reionisation algorithm.

We list our galaxy parameters and fiducial values in Table \ref{tab:galparams}.  These fiducial values are used throughout this paper except where otherwise indicated. Coefficients represented as numbers in the above equations are currently fixed in our model.

\cmfast\ includes an optional prescription for star formation inside molecularly cooled galaxies (MCG; c.f. \citealt{Qin20}), which could have an independent SHMR:
\begin{multline}\label{eq:mcgstar}
    \mu_{*,\mathrm{mcg}}(M_\mathrm{halo}) = f_{*,7} \left(
    \frac{M_h}{10^7 M_\odot} \right)^{\alpha_{*,\mathrm{mcg}}}
    \exp \left(\frac{-M_h}{M_\mathrm{cool}} \right) \\
    \exp \left(\frac{-M_\mathrm{LW}}{M_h} \right)
    M_h \frac{\Omega_b}{\Omega_m}
\end{multline}
MCGs are allowed to have different escape fractions when calculating the ionization field, as well as different spectra based on population III stellar models. The atomic cooling threshold mass $M_\mathrm{cool}$ acts as an upper turnover mass which separates this population from the atomically cooled galaxies (ACG's), and $M_\mathrm{LW}$ is the lower turnover mass, set by inhomogeneous Lyman-Werner radiation feedback and the relative velocities between dark matter and baryons. When the minihalo component is enabled by the user, the discrete halo model samples the MCG component from a log-normal distribution of the same width $\sigma_*$, but with a mean set by equation \ref{eq:mcgstar}, analogously to the ACG component. The specific star formation rates for MCGs are sampled from the same distributions as for the ACGs. Metallicities and X-ray luminosities are based on the combined stellar mass and star formation rate of both ACGs and MCGs. While we do not include MCGs in this work, they are available within \cmfastvfour\ with the full functionality implemented initially by \citet{Qin20} and expanded upon in \citet{Munoz22}.

To demonstrate the flexibility of our model, in Figure \ref{fig:scaling} we show two different parameter combinations capable of capturing the SHMR and SFMS seen in two different hydrodynamical cosmological simulations: the zoom-in suite SERRA \citep{Pallottini22} (setting $f_{*,10} = 0.16$, $\alpha_* = 0.1$, $t_* = 0.13$) and the large cosmological simulation Astrid \citep{Bird22} (setting $f_{*,10} = 0.005$, $\alpha_* = 0.65$, $t_* = 0.13$). While the simulations are roughly consistent at the highest halo masses, they diverge significantly at low halo masses, where there are no direct observations to anchor their sub-grid prescriptions. Our prescription is able to characterize such diverse distributions of galaxy properties, computing the corresponding signatures their radiation imprints in the IGM.  We illustrate such an approach in more detail in Section \ref{sec:forward}.

\begin{figure}
\includegraphics[width=\linewidth]{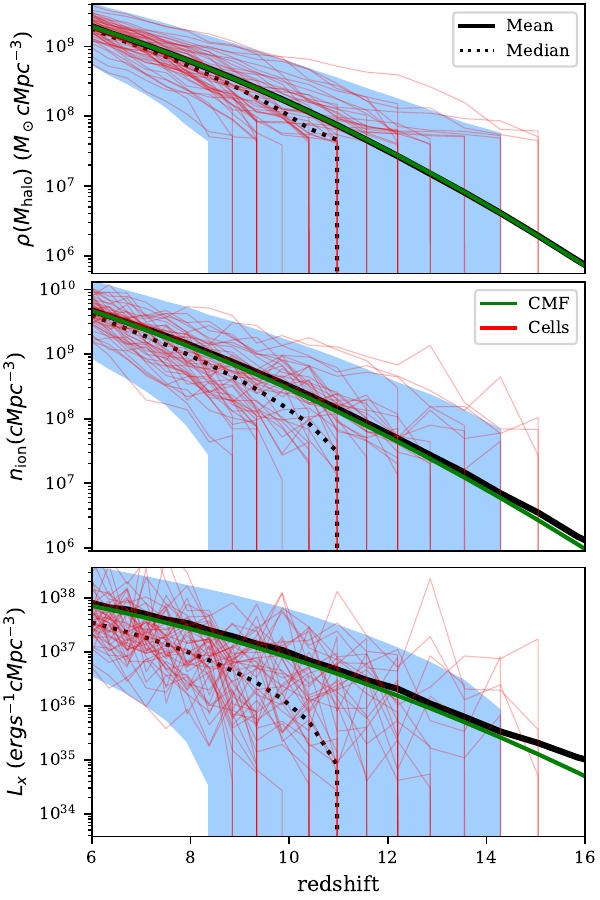}
    \caption{Galaxy properties within $32^3$ Lagrangian cells with side length 2 Mpc at mean density $(\delta=0$), including stochastically sampled halos above $10^8 M\odot$. We show halo mass density (top), cumulative ionising photon number (middle) and X-ray soft-band emissivity (bottom). Solid and dotted black lines show the mean and median of the distribution respectively, and green lines show the expected mean obtained by integrating the CHMF. blue shaded regions span the $[2.5,97.5]$ percentile range, and the thin red lines show 64 randomly selected individual cells. Changes over time in the summed galaxy properties are caused by the growth of halos, as well as the variability in the stellar to halo mass ratio and specific star formation rate.}
    \label{fig:histories}
\end{figure}

\section{Calculating radiation fields}\label{sec:radiation}
The galaxy sampling method described in sections \ref{sec:halosampling} and \ref{sec:galaxies} forms a new source model in \cmfastvfour.  Full halo mass histories are generated prior to calculating any radiation fields, moving backward in time in a user-defined redshift range. Each halo is moved from Lagrangian to Eulerian space, together with the matter field, using 2LPT.   We then grid the galaxy source field, starting from the highest redshifts, computing inhomogeneous radiation fields and taking into account feedback from previous time-steps \citep{Sobacchi13,Qin20,Munoz22}.

\begin{figure*}
    \centering
    \includegraphics[width=\linewidth]{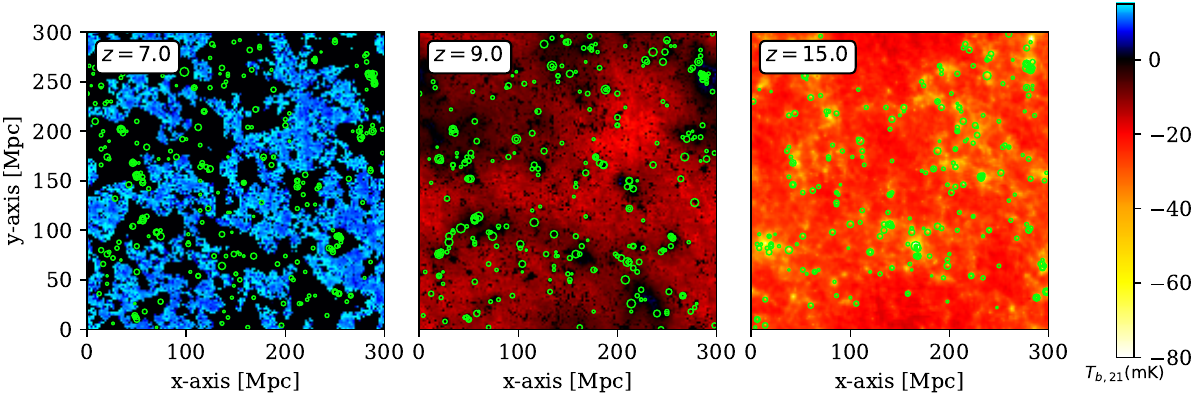}
    \caption{Co-location of galaxies and the 21cm brightness temperature fields in our illustrative simulation during key stages in the history of the CD and EoR.  We show slices at $z=14.8$, $z=9.0$, and $z=7.0$, corresponding to epochs of Lyman alpha coupling, X-ray heating, and reionization, respectively. Green circles denote 200 galaxies with the brightest UV magnitudes in each slice, with larger circles corresponding to brighter galaxies.}
    \label{fig:colocation}
\end{figure*}

The X-ray and Lyman alpha specific intensities ($J_X$, $J_\alpha$) at redshift $z$ and frequency $\nu$ is computed by integrating back along each cell's lightcone:
\begin{multline}\label{eq:heatintegral}
    J_i(\mathbf{x},z,\nu) = \frac{(1+z)^3}{4\pi}  \int_{z}^{\infty} dz' \frac{cdt}{dz'} \sum_{x'} W_\mathrm{ts}(x',z,z') \\
    \epsilon_i(\mathbf{x'},z',\nu') e^{-\tau_i(z,z')}
\end{multline}
where $\mathbf{x}$ is the cell's location, $\epsilon_i$ is the specific comoving emissivity at a location $\mathbf{x'}$, redshift $z'$ and emitted frequency $\nu' = \nu (1+z)/(1+z')$. The optical depth, $\tau_i$, accounts for attenuation due to the intervening IGM. $W_\mathrm{ts}(x',z,z')$ is a window function which selects sources at the correct distance from $\mathbf{x}$, given $z$ and $z'$. Heating and ionisation rates are then computed by integrating $J_i$ over $\nu$, and relevant cross sections. Finally, the residual ionisation by X-rays, kinetic temperature, and spin temperature of the mostly neutral hydrogen are determined from these rates.
Further details on these models, including how emissivities $\epsilon_i$ are calculated from star formation rates of ACGs and MCGs,  and how IGM attenuation is computed can be found in \citet{Mesinger11,Sobacchi13,Qin20} and \citet{Munoz22}.

In the previous versions of {\tt 21cmFAST}, when computing the emissivity at a comoving distance $R(z, z')$, the Eulerian overdensity was evolved back from $z$ via the ratio of linear growth factors, $D(z')/D(z)$ and the window function $W_\mathrm{ts}(x',z,z')$ was a spherical top-hat. This was done to speed up the calculation as both the absorbing IGM and emitting galaxies were calculated directly from density fields pre-filtered with the same window functions. With the discrete source field in {\tt 21cmFASTv4}, we can directly use the previous inhomogeneous emissivity fields $\epsilon_i(\vec{x},z'(R))$ and a spherical shell filter of finite width.
\begin{equation}\label{eq:annularfilter}
    W_\mathrm{ts}(r,R_\mathrm{o},R_\mathrm{i}) = 
    \begin{cases}
        \frac{3}{4 \pi (R_o^3-R_i^3)} & \text{if } R_\mathrm{i} < r < R_\mathrm{o}\\
        0  & \text{otherwise }
    \end{cases}
\end{equation}
where $R_\mathrm{i}$ and $R_\mathrm{o}$ are the inner and outer comoving radii of the shell respectively\footnote{The numerical (trapezoidal) integration of equation \ref{eq:heatintegral} forces us to use a finite $\Delta R \equiv R_\mathrm{o} - R_\mathrm{i}$,  We confirm that if we use the infinitely thin spherical shell limit window function (i.e. $\Delta R\rightarrow 0$), we obtain noticeable aliasing and miss small scale emissivity fluctuations.}. For computational efficiency, we apply this filter to our gridded emissivities in Fourier space, using its Fourier transform:
\begin{multline}\label{eq:annularfilterFT}
    W_\mathrm{ts}(k,R_\mathrm{i},R_\mathrm{o}) = \frac{3}{k^3(R_\mathrm{o}^3 - R_\mathrm{i}^3)}
    \{ [\sin(kR_\mathrm{i}) - kR_\mathrm{i} \cos(kR_\mathrm{i})] \\- [\sin(kR_\mathrm{o}) - kR_\mathrm{o} \cos(kR_\mathrm{o})] \}
\end{multline}

Because the mean free path of Lyman limit photons through the neutral IGM is very short, reionization is a fairly local process with sharp discontinuities along ionization fronts, and is therefore not well captured by the lightcone integration of eq (\ref{eq:heatintegral}). Instead reionization is calculated via the excursion set algorithm \citep{Furlanetto04,Zahn11,Mesinger11}, where the source field is filtered on a series of decreasing radii, marking as ionized those cells that receive more ionizing photons than neutral atoms\footnote{We make the standard assumption that helium is singly ionized together with hydrogen.} plus recombinations:
\begin{equation}
    \sum_{x'} W_\mathrm{ion}(r,R) n_\mathrm{ion}(\mathbf{x'}) > \frac{(1+\delta)\rho_\mathrm{crit}}{m_p} + n_{\mathrm{rec}}(\mathbf{x})
\end{equation}
where the left-hand side is the number of ionising photons reaching a cell, directly calculated from the cumulative ionising photon number $n_\mathrm{ion}$ (see equation \ref{eq:nion}) using a filter function $W_\mathrm{ion}(r,R)$ dependent on the distance between the source cell and absorber cell $r = |\mathbf{x'} - \mathbf{x}|$ and $n_{\mathrm{rec}}$ is the number of recombinations that have occurred within the cell.
In previous versions of the code we used a spherical top-hat or sharp-k filter for $W_\mathrm{ion}(r,R)$, with the maximum allowed scale set by the mean free path through the ionized medium, $\lambda_{\rm lls}$. Since we use a discrete ionizing source field, here we implement the exponential window function suggested by \citet{Davies22} (their MFP -- $\epsilon (r)$ method):
\begin{equation}
    W_\mathrm{ion}(r,R,\lambda) = 
    \begin{cases}
        \frac{3}{4 \pi R^3}e^{-r/\lambda_{\rm lls}} & \text{if } r < R\\
        0  & \text{if } r \geq R
    \end{cases}
\end{equation}
where we take $\lambda_{\rm lls} = 25 \ \mathrm{cMpc}/h$, based on lyman limit system measurements at $z\lesssim5$ \citep{Songaila10}.  This parameter can also be varied, though for simplicity we take a fixed value in this work\footnote{The EoR history is insensitive to this choice when accounting for inhomogeneous sub-grid recombinations (e.g. \citealt{Qin24inference}).  This is because the total mean free path during the majority of the EoR $\lambda_{\rm tot}$ is dominated by the typical distance to HI patches, $\lambda_{\rm EoR}$, with $\lambda_{\rm tot}^{-1} = \lambda_{\rm EoR}^{-1} + \lambda_{\rm lls}^{-1}$ (e.g. \citealt{Alvarez12}).  Indeed,  recent measurements \citep{Becker21} suggest a much shorter mean-free path $< 10 \mathrm{cMpc}$ for ionising photons at $z\sim6$, which could be due in part to the contribution of residual HI patches and/or details in how it is computed from observations \citep{Satyavolu24, Qin24inference}. }.


Once the thermal and ionisation state of the gas is determined by the above models, we compute the 21cm signal in the form of the differential brightness temperature $dT_{b}$ (e.g. \citealt{Furlanetto06})
\begin{multline}\label{21cmbt}
    dT_{b} = 27 \left( \frac{\Omega_bh^2}{0.023} \right) \left( \frac{0.15}{\Omega_mh^2} \right)^{0.5} \\
    \left( \frac{1+z}{10} \right)^{0.5} \left( 1 -  \frac{T_\gamma}{T_s} \right) \left( \frac{H(z)}{\frac{dv_r}{dr} + H(z)} \right)
     x_\mathrm{HI} (1+\delta) 
\end{multline}
where $T_\gamma$ is the temperature of the cosmic microwave background and each cell is described by its overdensity $\delta$, neutral fraction $x_\mathrm{HI}$, spin temperature $T_s$ and line of sight velocity gradient $\frac{dv_r}{dr}$. $dT_b$ quantifies the difference in intensities of the IGM against the cosmic microwave background at the rest frame wavelength of 21cm.

\begin{figure*}
\centering
\includegraphics[width=0.85\textwidth,height=\textheight,keepaspectratio]{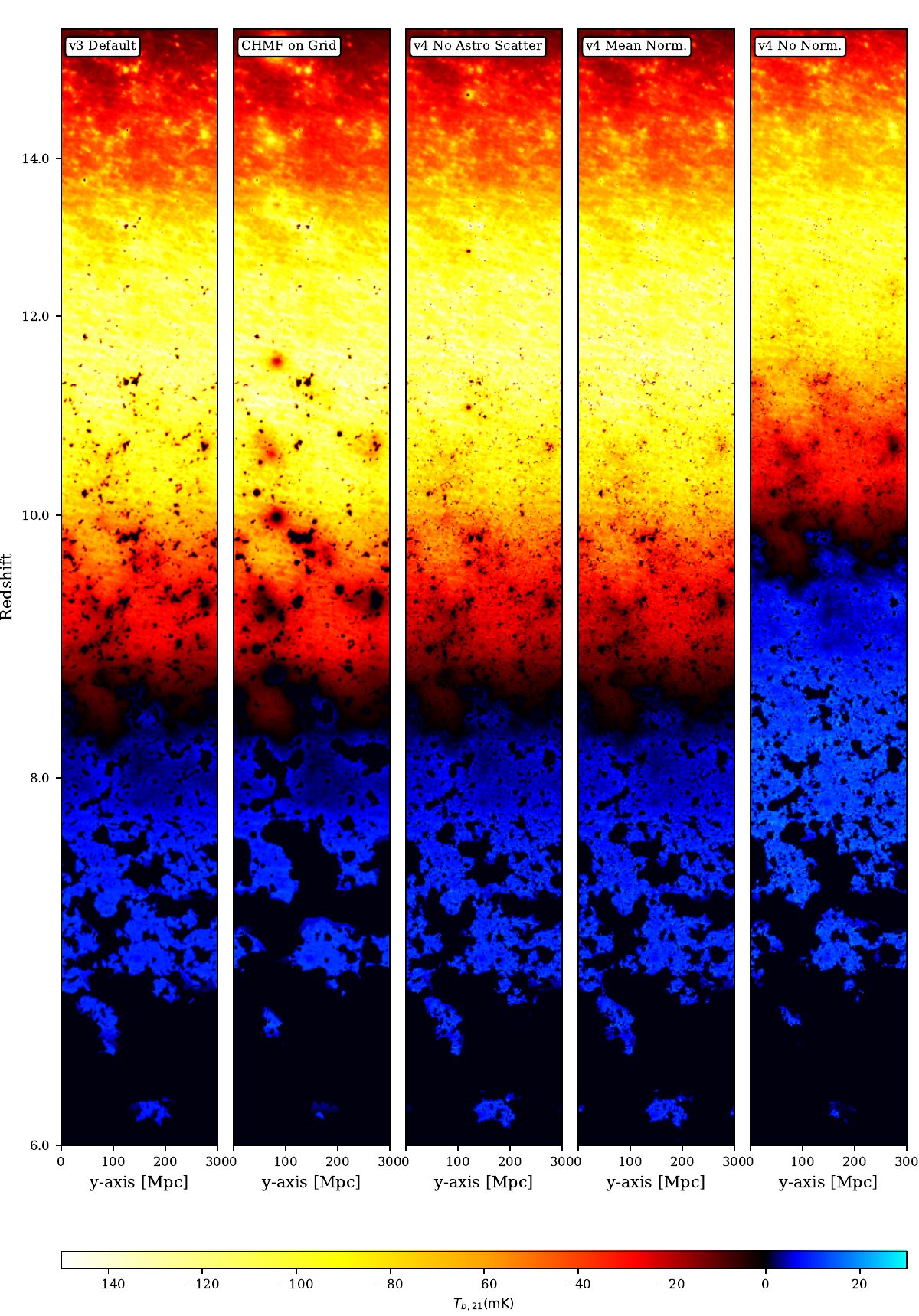}
    \caption{21cm brightness temperature lightcones from four runs of \cmfast. From left to right: ``v3 Default" shows the previous default model, where the source field is calculated on the filtered density grids, ``CHMF on Grid" shows a model where there is a one-to-one relationship between cell density and emissivities, based on the integral of the CHMF. ``v4 No Astro Scatter" shows the effects of our halo sampler described in section \ref{sec:halosampling}, without any stochasticity in the galaxy properties. We show two cases for the model with stochasticity in the halo mass distribution and galaxy properties, which differ in their interpretations of the scaling relations. ``v4 Mean Norm." shows the case where the scaling relations are defined by the mean of each property, and ``v4 No Norm." shows the case where the scaling relations are defined by the mean logarithm.  In the latter case, increasing the log-normal scatter also increases the mean emissivities at all wavelengths.}
    \label{fig:compare_lc}
\end{figure*}

\section{Illustrative examples}\label{sec:example}
\begin{table*}
    \centering
    \begin{tabular}{ c|c|c|c|c }
        \hline
        Name & Sources Defined on Grid & Stochastic Halos & Astrophysical Scatter & Scaling Relation \\
        \hline
        v3 Default & \rcross & \rcross & \rcross & N/A \\
        CHMF on Grid & \gcheck & \rcross & \rcross & N/A \\
        v4 No Astro Scatter & \gcheck & \gcheck & \rcross & N/A \\
        v4 Mean Normalisation & \gcheck & \gcheck & \gcheck & fixed mean \\
        v4 No Mean Norm. & \gcheck & \gcheck & \gcheck & fixed mean log \\ 
    \end{tabular}
    \caption{Settings of the simulations performed to investigate the effect of stochasticity in halo masses and bulk galaxy properties.}
    \label{tab:stocruns}
\end{table*}

To showcase some novelties of this model, we perform a simulation using the halo sampler described in section \ref{sec:halosampling}, galaxy property distributions described in section \ref{sec:galaxies}, and EoR/CD model described in section \ref{sec:radiation}. We simulate a box with side-length $300 \mathrm{Mpc}$ from $z=35$ to $z=5$. Initial conditions are sampled on a $600^3$ grid, which is downsampled to $150^3$ for the purposes of sampling halos as well as performing the 2LPT and IGM calculations. All astrophysical parameters for this run are laid out in table \ref{tab:galparams}. We integrate the average CHMF in the range $10^7M_\odot < M_h < 10^8M_\odot$, and perform stochastic sampling above $10^8 M_\odot$.

In Figure \ref{fig:histories} we show the redshift evolution of the volume densities of the total halo mass, cumulative number of ionizing photons per unit volume, and the rest-frame soft-band X-ray emissivity for $32^3$ Lagrangian cells of side length $2 \mathrm{Mpc}$ at cosmic mean density. Despite having the same density by construction, each cell can vary greatly in its properties as new galaxies form and experience periods of high or low star formation.

We show how the 21cm signal is affected by the spatial distribution of galaxies at three key epochs in figure \ref{fig:colocation}. At each of these epochs, the changes in the IGM are being driven by cosmic radiation fields sourced by galaxies (see, e.g. \citealt{Furlanetto06}). At $z=14.8$ the 21cm cold spots are close to the brightest galaxies, as Lyman alpha radiation couples the spin temperature to the  kinetic temperature which is below the CMB temperature at the time. Around $z=9.0$ areas close to galaxy overdensities have a higher 21cm brightness temperature, as X-rays from nearby sources heat the IGM. Finally during reionisation at $z=7$, the coldest regions are again near galaxy overdensities, as they have ionised their surroundings resulting in a brightness temperature close to zero.  Although the cosmic radiation are dominated by the more abundant, faint galaxies not visible in the figure, they cluster around the brightest galaxies denoted with green circles, allowing us to see these spatial correlations between bright galaxies and IGM properties.

\section{Impact of stochasticity on the 21cm signal}\label{sec:comparison}
Our new simulations allow us to quantify the importance of stochasticity in both the halo field as well as the halo -- galaxy connection.  
To do this, we run five iterations of \cmfastvfour, all using the same Sheth-Tormen halo mass function and a box side-length of $300 \, \mathrm{Mpc}$. Initial conditions are sampled on a $600^3$ grid, while the galaxy, IGM and radiation fields are computed on $200^3$ grids. Astrophysical parameters are as presented in Table \ref{tab:galparams}, with two exceptions: the mean X-ray luminosity at fixed star formation rate, which is set to a constant $L_X / \mathrm{SFR} = 10^{40} \mathrm{erg s^{-1} M_\odot^{-1} yr}$, and the SHMR is confined to a single power-law, effectively setting $\alpha_{*2} = \alpha_* = 0.5$. Additionally, we switch off the mean-free path filter described in section \ref{sec:radiation} and \citet{Davies22}, using a spherical top-hat ionisation filter for all simulations. These changes were made for easier comparison with \cmfast v3 and to highlight the effects of the stochastic halo field.

\begin{figure}
	\includegraphics[width=\linewidth,keepaspectratio]{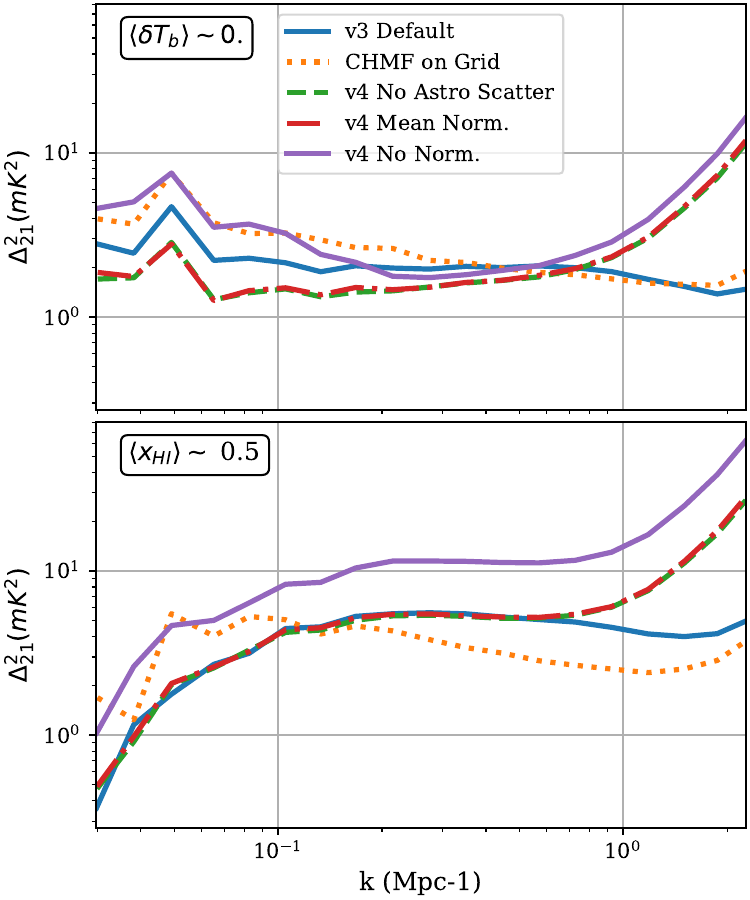}
    \caption{21cm power spectra shown for the same five models as in Fig. \ref{fig:compare_lc}. We show the power-spectra at two points in cosmic history: the midway point of the EoR (the snapshot at which the total ionised fraction is closest to 50\%), as well during X-ray heating where the volume-weighted mean brightness temperature $\langle dT_{b,21} \rangle$ is approximately zero.}
    \label{fig:powerspec}
\end{figure}

\begin{figure*}
    \includegraphics[width=\textwidth,keepaspectratio]{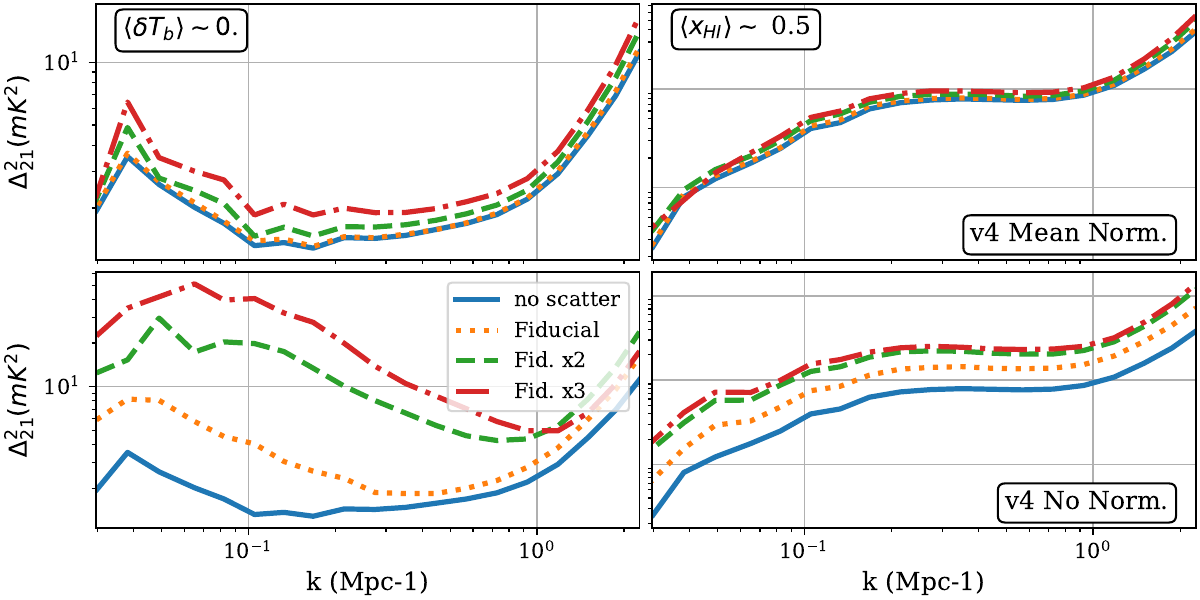}
    \caption{21cm brightness temperature power spectra from {\tt 21cmFASTv4} runs with differing levels of astrophysical scatter. The top rows show the runs where the scaling relations have a fixed (linear) mean as scatter increases, whereas the bottom rows show runs where the mean is allowed to increase with the scatter (i.e. at a fixed log-log mean relation).  The left column shows the epoch during X-ray heating where $\langle dT_{b,21} \rangle = 0$ mK, and the right column shows the midpoint of reionisation $\langle x_{HI} \rangle = 0.5$. The 21cm power spectra for the mean normalised runs show a much milder evolution with astrophysical scatter, however the scatter still affects the power spectra when boosted 2 to 3 times our fiducial, as might be needed to explain recent {\it JWST} UV LF observations.
    }
    \label{fig:boost_ps}
\end{figure*}

The first iteration titled ``v3 Default" shows the previous default source model, in which the galaxies are not localized but instead mean abundances are calculated from CHMFs in expanding spherical volumes centred on each cell. The second run titled ``CHMF on Grid" has a fixed, one-to-one relationship between the source grid and the evolved density grid based on the integral of the conditional mass function on the scale of a grid cell. The emissivity fields produced from this are then filtered in the same way as the sampled halo fields \footnote{This model is analogous to the source models labelled FFRT-P in \citet{Davies22} or ESF-E in \citet{Trac22}}. As in \cmfast v3, both the ``CHMF on Grid" and ``v3 Default" models have their global mean emissivities at each wavelength fixed to the expected mean, which is obtained by integrating the unconditional mass functions at each redshift. The third iteration titled ``v4 no astro scatter" uses our halo sampler described in section \ref{sec:halosampling}, however with no log-normal scatter around the galaxy scaling relations described in section \ref{sec:galaxies} (by setting $\sigma_* = \sigma_\mathrm{SFR,lim} = \sigma_\mathrm{SFR,idx} = \sigma_\mathrm{X} = 0$). The fourth and fifth iterations ``v4 Mean Norm" and ``v4 No Mean Norm" both use the halo sampler with log-normal scatter in galaxy properties. The key difference between these two models is whether the scaling relations (equations \ref{eq:acgstars}, \ref{eq:sfrmean}, \ref{eq:xraymean}, and \ref{eq:mcgstar}) are defined to have a fixed mean as the scatter increases \footnote{Because we assume log-normal scatter, the only difference between these models is a factor of $10^{\sigma_{\mathrm{\log x}}^2/2}$ in each scaling relation, where the "normalised" scaling relations represent a mean $\mu_x$, as opposed to the median (or equivalently the exponent of the mean logarithm $10^{\mu_{\log x}}$) of the galaxy property (see section \ref{sec:galaxies}).  These conditional probability distributions need not be log-normal, in which case the normalization to obtain a fixed global mean emissivity could be computed numerically.}. In ``v4 No Mean Norm" the mean log-log scaling relation is kept fixed, and therefore as log-normal scatter is increased, the linear mean increases (c.f.  Appendix B in \citet{Nikolic24}).  In ``v4 Mean Norm" on the other hand, the linear mean is kept constant as the scatter is changed.

The key differences between each model are summarised in Table \ref{tab:stocruns}. The 21cm brightness temperature lightcones are shown in Figure \ref{fig:compare_lc}.  In Fig. \ref{fig:powerspec} we plot the spherically-averaged 21cm power spectra $\Delta^2_{21}(k,z) =  \frac{k^3}{2\pi^2 V}\langle |dT_{b,21}(k,z) - \langle dT_{b,21} \rangle(z)|^2 \rangle_k$ at two epochs: the midpoint of the EoR ($\langle x_{\rm HI} \rangle \approx 0.5$) is shown in the bottom panel while the epoch of heating when the mean brightness temperature is zero ($\langle dT_{b,21} \rangle \approx 0$) is shown in the top panel.

Comparing the ``v3 Default" and ``CHMF on Grid" outputs highlights the impact of defining the source model only at the cell scale, rather than over a range of larger spheres. Indeed ``CHMF on Grid" is a notable outlier between our various test cases (aside from ``v4 No Mean Norm." which is discussed below), having patchier heating and ionization morphologies characterized by larger ionized and heated regions.    Applying CHMFs only on relatively small Lagrangian scales corresponding to the simulation grid would miss the most massive halos (e.g. \citealt{Barkana04}; Appendix A in \citealt{Nikolic24}).  Attempting to correct for this by instead conditioning the HMFs on the evolved (Eulerian) cell densities (i.e. "CHMF on Grid"), overcompensates and instead results in spuriously massive halos in the densest cells (e.g. footnote 4 in \citealt{Sobacchi14}; Fig 13 in \citealt{Trac22}).  This results in ``CHMF on Grid" overestimating the 21cm large scale power, as seen in Fig. \ref{fig:powerspec}.

Comparing the ``v3 Default" and ``v4 No Astro Scatter" illustrates the importance of having a stochastic halo finder in the source model.  Understandably, having discrete sources results in small-scale "shot-noise" in the power spectra, as seen in Fig. \ref{fig:powerspec} at $k\gtrsim1$Mpc$^{-1}$.  Differences on larger scales and from visual inspection of the lightcones are modest.

Comparing instead ``v4 No Astro Scatter" to ``v4 Mean Normalisation" isolates the importance of astrophysical scatter in our default model, {\it at fixed mean emissivities}.  We see that when the emissivities are adjusted to have the same mean, the scatter in galaxy properties has a negligible impact on the 21cm signal.  This is due to the fact that variations in halo mass distribution cause a much wider scatter $(> 1 \mathrm{dex})$ in emissivities at fixed matter overdensity, such that they dominate when added in quadrature to the scatters in the galaxy properties $\leq 0.5 \mathrm{dex}$. 
The relative importance of scatter in halo abundances vs astrophysics will depend however on the details of the model and sampling method, and any choice that reduces the variance in the halo mass distribution (such as the stricter halo mass conservation criteria in \citealt{Nikolic24} which limits the total halo mass variation within a $5 \mathrm{Mpc}$ region to $\pm 10\%$) will increase the relative importance of astrophysical scatter.

However, if including scatter around log-normal galaxy scaling relations, without adjusting the means of the distribution, the resulting impact is significant.  We can see this comparing ``v4 Mean Normalisation" with ``v4 No Normalisation".  The milestones of the 21cm signal in the later model are shifted to earlier times, because of the shift in the mean when widening a log-normal distribution. 
The most dramatic shift, resulting in a large increase in power (c.f. Fig. \ref{fig:powerspec}), occurs during the X-ray heating epoch.  Compared to the UV emissivity,  the X-ray emissivity is sensitive to an additional source of scatter coming from the $P(L_X | \mathrm{SFR}, Z)$ distribution.  This serves as a caution against using empirically-derived mean log-log scaling relations directly in theoretical models without accounting for scatter around the mean (see also \citealt{Nikolic24}).

\subsection{Varying the level of astrophysical scatter}

As mentioned before, the astrophysics of EoR/CD galaxies is very poorly known, and our conclusions in the previous section could depend on our choice of fiducial parameters.  Our fiducial parameters characterizing the halo--galaxy connection were loosely motivated by hydrodynamic cosmological simulations.  However, output from different simulations can disagree significantly (e.g. the two simulations shown in Fig. 5 have mean SHMR that are different by up to two orders of magnitude).  Moreover, recent observations of UV luminosity functions at $z>11$ using {\it JWST} are difficult to explain with current simulations and might require significantly more scatter in galaxy properties (e.g. \citealt{Pallottini23,Gelli24,Nikolic24}).

Here we show how increasing (or decreasing) the level of astrophysical scatter impacts observables.  We perform runs with two or three times the scatter of our fiducial runs, where a $\times Y$ boost corresponds to substituting $\sigma_{\log x} \rightarrow \sigma_{\log x} + \log Y$ in each of our scaling relations. As with the previous simulations, we include runs where the galaxy scaling relations define a fixed mean, and those where the mean is allowed to increase with the log-normal scatter. The power spectra from these runs at the same two epochs as in Figure \ref{fig:powerspec} are shown in Figure \ref{fig:boost_ps}.

In the case where the mean of the galaxy property distributions are not fixed, there is significant change in the power spectrum at all epochs driven by an increase in the mean stellar mass, star formation rate and X-ray luminosity. This mean shift causes reionisation and reheating to progress faster, resulting in a higher power on all scales simply due to the higher redshift. In the case where the galaxy scaling relations define a fixed mean, we begin to see a $\sim 10\%$ increase in 21cm power during X-ray heating at $\gtrsim 2 \times$ fiducial scatter, and a $\sim 50\%$ increase for $3 \times$ fiducial scatter. At these increased levels, the astrophysical scatter begins to dominate over the scatter introduced by the halo sampling, and the 21cm signal can be used to constrain the width of galaxy property distributions.

\begin{figure}
\includegraphics[width=\linewidth,keepaspectratio]{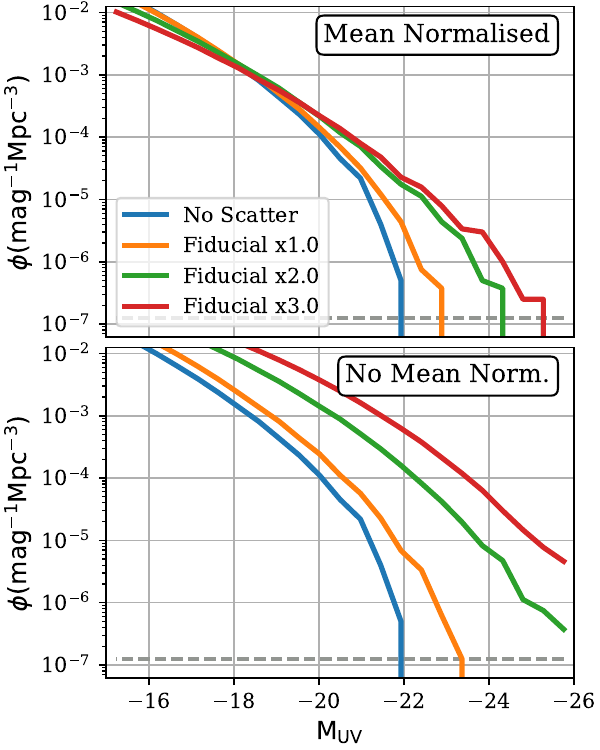}
    \caption{The UV luminosity functions at $z=8$  by setting the scatter in the galaxy properties to either zero (blue line), equal to the fiducial values from Table \ref{tab:galparams} (orange line), and increased by factors of 2 (green line) or 3 (red line). The top panel shows runs where the scaling relations are normalised to have the same linear mean, and the bottom panel shows runs where log-normal scatter is simply added to a fixed log-log mean relation. In both cases increasing scatter results in a flattening of the relation, however in the later case it also results in significantly brighter galaxies overall.}
    \label{fig:uvlf}
\end{figure}

Although our fiducial level of scatter has little effect on the 21cm power spectrum, it leaves a significant imprint on the UV luminosity function, as shown in Figure \ref{fig:uvlf}. Simply adding lognormal scatter to a scaling relation without keeping the means fixed drives up the UV luminosity function at all magnitudes. When normalised to have the same mean galaxy properties, the UVLF flattens as scatter increases, resulting in a factor of 10 increase in the abundance of bright galaxies $M_{UV} \sim -22$. By utilising observables from both the IGM and galaxies, we will be able to constrain the normalisation and scatter in bulk galaxy property distributions.

\section{Forward modelling multi-tracer data}\label{sec:forward}
The halo sampler described in Sect. \ref{sec:halosampling} combined with the flexible halo--galaxy connection described in Sect. \ref{sec:galaxies} form a new optional source model in \cmfastvfour.  In this set-up, multi-wavelength radiation sources are explicitly localized, and have their properties sampled self-consistently according to given distributions.  Any additional quantities such as line luminosities can be easily calculated in post-processing from either the gridded galaxy fields or catalogues.  Our simulations can thus self-consistently compute lightcones of galaxies, their radiation fields and the corresponding IGM evolution.  These can be used to forward model IGM observations, galaxy surveys, LIMs, as well as their cross-correlation allowing us to {\it infer cosmological parameters or those characterizing the halo--galaxy connection} (e.g. Fig. \ref{fig:fmschematic}). 

\begin{figure*}
\centering
\includegraphics[width=\textwidth,height=\textheight,keepaspectratio]{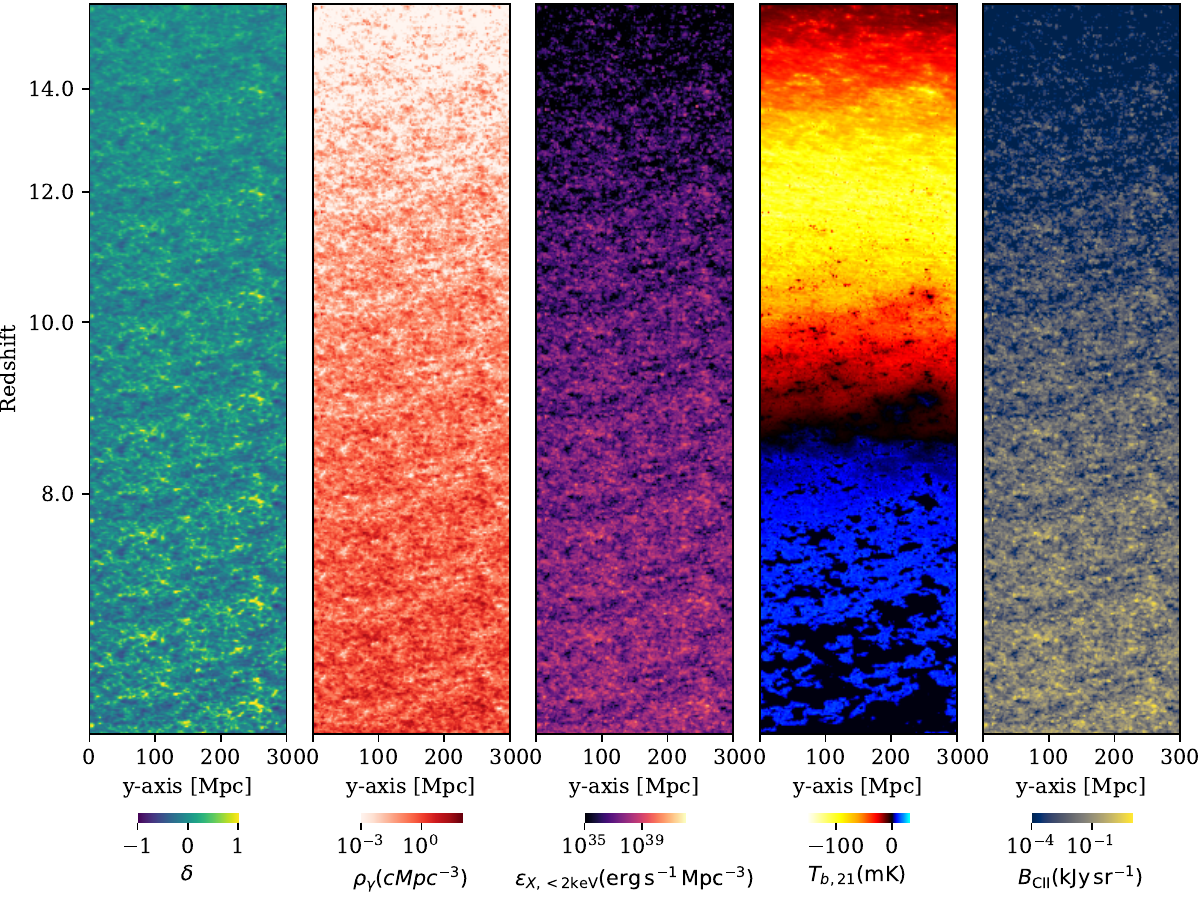}
    \caption{Lightcones containing example outputs from  \cmfastvfour\ with default parameters. The leftmost panel shows the evolved matter density field. The next two panels show 
    gridded galaxy properties: cumulative ionising photon number and soft-band X-ray emissivity. The two rightmost panels show observables: the 21cm brightness temperature and CII surface brightness density.}
    \label{fig:halo_lc}
\end{figure*}

We show full lightcones of overdensity, cumulative ionising photon density, soft-band X-ray emissivity, 21cm brightness temperature, and CII surface brightness density from \cmfastvfour\ with our fiducial parameters (Table \ref{tab:galparams}) in Figure \ref{fig:halo_lc}. We calculate the latter using the following simple linear relation \citep{DeLooze14} for CII luminosity for all galaxies:
\begin{equation}\label{eq:ciilum}
    \log \left( \frac{L_\mathrm{CII}}{L_\odot} \right) = 7.06 + \log \left( \frac{\mathrm{SFR}}{M_\odot yr^{-1}} \right)
\end{equation}
We sum all sources within each cell to obtain the final surface brightness density, for simplicity assuming all galaxies are at the centre of the cell
\begin{equation}
    B_\mathrm{CII} = \frac{1}{1.04 \times 10^{-3} D_L^2 \Omega \Delta \nu} \sum_{\mathrm{cell}} L_\mathrm{CII} ~ ,
\end{equation}
where $\Delta \nu$ is the frequency width of the cell in GHz, $\Omega$ is the angular size of the cell and $D_L$ is the luminosity distance in Mpc.

\begin{figure*}
\centering
\includegraphics[width=0.85\textwidth,keepaspectratio]{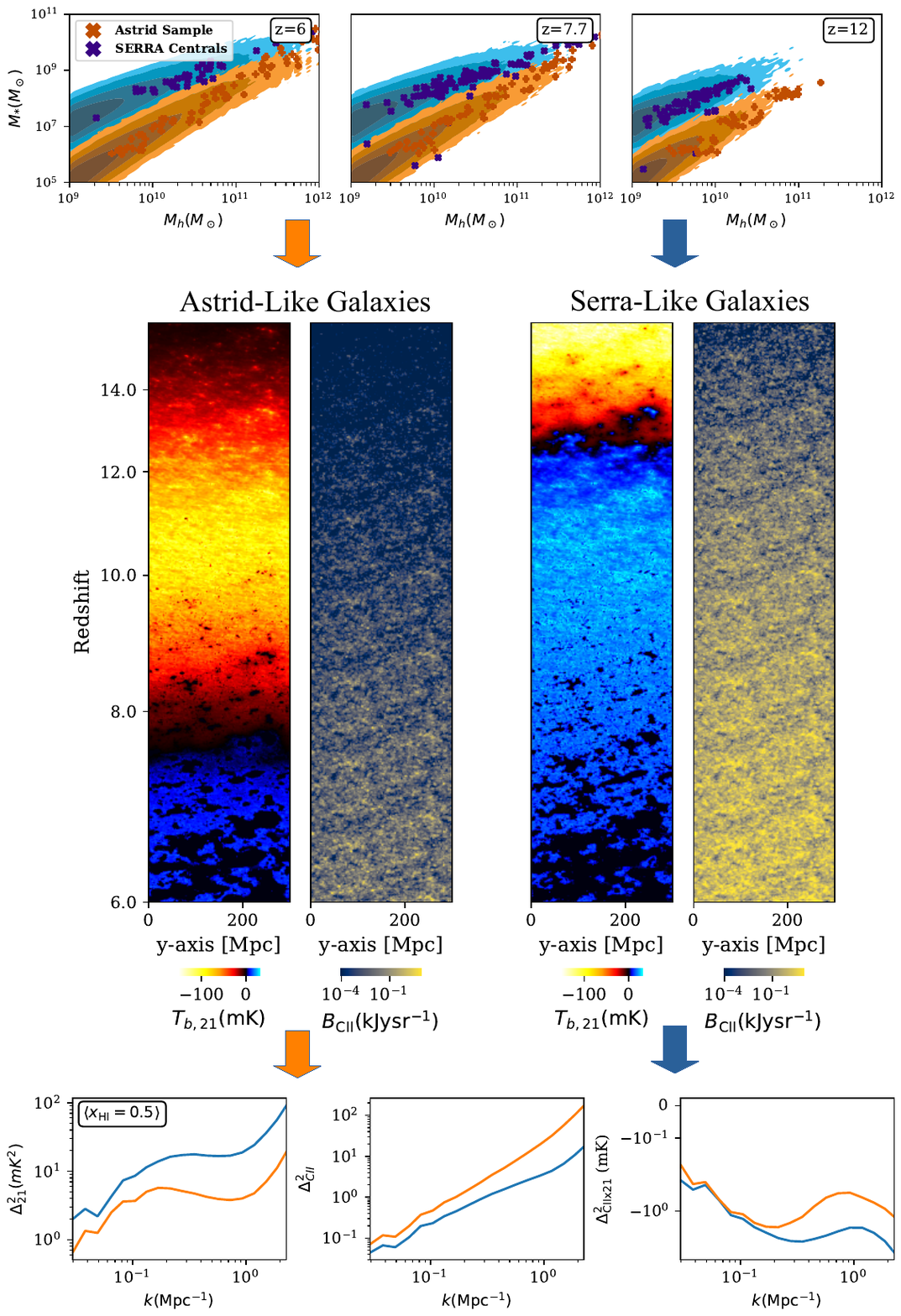}
    \caption{Schematic showing how different galaxy populations affect IGM and LIM observables. On the left we show the SHMR panels from figure \ref{fig:scaling}, where we set our galaxy scaling relations are chosen to be consistent with either the SERRA or Astrid cosmological simulations. The middle panels show the resulting 21cm brightness temperature and CII surface brightness density lightcones in the range $6<z<16$. The right panels show the auto power spectra from both fields, as well as their cross power spectrum at the midpoint of reionisation $\langle x_\mathrm{HI} \rangle =0.5$. Each galaxy population produces a starkly different cosmic dawn scenario, easily distinguished by the power spectra of the 21cm and/or CII fields.}
    \label{fig:2parameters}
\end{figure*}

As a simple demonstration of such a multi-tracer approach, we run two simulations with the same dimensions, mass function, and sampling algorithm as those in sections \ref{sec:example} and \ref{sec:comparison} but with two sets of astrophysical parameters chosen to characterize simulated galaxies from SERRA \citep{Pallottini22} and Astrid \citep{Bird22}. As discussed in section \ref{sec:galaxies}, this is done by setting the parameters related to the SHMR and SFMS to be consistent with the distributions from each simulation. In the case of SERRA we use $f_{*,10} = 0.16$, $\alpha_* = 0.1$, and $t_* = 0.13$, and in the case of Astrid we use $f_{*,10} = 0.005$, $\alpha_* = 0.65$, and $t_* = 0.13$. Both runs have their escape fractions tuned so that reionisation finishes around $z=5.5$, and all other parameters are set according to Table \ref{tab:galparams}. By construction, both models are consistent with current galaxy observations.  However, they imply very different star formation efficiencies in the abundant faint galaxies below direct detection limits.  Thus, these two galaxy formation scenarios would result in very different cosmic radiation fields, and IGM and/or LIM observations could be used to distinguish between them. We illustrate this in Fig. \ref{fig:2parameters}  where we compute the 21cm brightness temperature fields and CII line intensity maps corresponding to these two models. 

The vastly different stellar populations between these models are easily distinguished in the X-ray heating history, with the more efficient star formation of SERRA suggesting the Epoch of Heating occurs at $z\sim13$, while our Astrid-like galaxies imply it occurs at $z\sim8$.

Even with a similar reionisation timing, we can clearly distinguish these scenarios in the auto and cross power spectra of 21cm brightness temperature and CII surface brightness density.
\begin{equation}
     \Delta^2_{21}(k,z) =  \frac{k^3}{2\pi^2 V}\langle |dT_{b,21}(k,z) - \langle dT_{b,21} \rangle(z)|^2 \rangle_k
\end{equation}
\begin{equation}
     \Delta^2_\mathrm{CII}(k,z) =  \frac{k^3}{2\pi^2 V}\langle \delta_\mathrm{CII}^2 \rangle_k
\end{equation}
\begin{equation}
     \Delta^2_\mathrm{CIIx21}(k,z) =  \frac{k^3}{2\pi^2 V}\langle \delta_\mathrm{CII} \left( dT_{b,21}(k,z) - \langle dT_{b,21} \rangle(z) \right) \rangle_k
\end{equation}
where $\delta_\mathrm{CII} = B_\mathrm{CII}(x,z)/\langle B_\mathrm{CII}\rangle(z) - 1$. The flatter stellar-to-halo mass relation in the run matched to SERRA means that less-massive halos have a more dominant contribution to cosmic radiation fields.
This results in smaller HII regions meaning higher power at large $k$ (e.g. \citealt{McQuinn07}), a less biased CII surface brightness density field, as well as a stronger anti-correlation between 21cm and CII on small scales.

This example shows the effect of changing parameters related to one of our galaxy scaling relations on two observable probes. When considering all of our scaling relations, it is important to note that some parameters controlling the galaxy properties can have degenerate effects on certain observables (for example, the SHMR and the escape fraction normalisations will have a degenerate effect on global reionisation history). These degeneracies must be broken by comparing our models to multiple observables at multiple redshifts. Our new framework allows us to perform inference on the global 21cm signal, 21cm power spectrum, galaxy luminosity functions, line intensity maps and radiative backgrounds simultaneously, including their cross-correlations. As a result we will be able to obtain much tighter constraints on the physics driving the first galaxies and their effects on the IGM.

\section{Conclusions}\label{sec:conclusion}
In this paper we have introduced a fast and flexible method for producing galaxy populations within the semi-numerical cosmological simulation code \cmfast.
Dark matter halos are identified using a combination of previous Lagrangian halo finding and a new, efficient merger-tree algorithm.
Galaxies are assigned to dark matter halos using flexible conditional probability distributions, motivated by well-established empirical relations.
Radiation fields are computed using these discrete galaxy source fields, including also the mean contribution of unresolved sources in each grid cell if the user selects a low halo mass resolution. The model can produce stochastic halo populations and associated EoR/CD history down to $10^8 M_\odot$ in a $300 \mathrm{Mpc}$ lightcone from $z=35$ to $z=6$ in approximately 3 core hours.  This can be reduced even further by raising the halo mass threshold for the sampler, and assuming mean emissivities for the unresolved galaxy component.


The model has parameters controlling the mean and scatter in the galaxy scaling relations, resulting in a wide range of possible scenarios for the co-evolution of galaxies and the IGM during cosmic dawn and reionisation. We show that our parametrization is sufficiently flexible to characterize the galaxy properties predicted by very different hydrodynamic simulations.  Additionally, \cmfastvfour\ is now able to output galaxy properties in the form of both halo catalogs and gridded halo properties. Statistics calculated from the galaxy field can be easily produced, and corresponding observations can be used in our inference pipelines, allowing us to tighten constraints on our model parameters and use cross-correlations to probe the galaxy-IGM connection.

We find that the stochastic source field produces significant shot-noise in the 21cm power spectrum at all redshifts. At our fiducial levels, the 21cm power-spectrum is unaffected by the scatter in the stellar-to-halo mass relation, star-forming main sequence, and X-ray luminosity, as they are dominated by stocasticity in the halo mass distribution. However we still see their effects in the UV luminosity function, which flattens as the scatter in the scaling relations increases. Additionally, increased levels of astrophysical scatter, which may be implied by recent \textit{JWST} observations, can be constrained by the 21cm power spectrum as they begin to overcome the scatter in halo mass at 2-3 times our fiducial levels. These observables are highly complementary; by comparing a wide range of observations with our model we will be able to extract the maximum information regarding the early universe from current and next-generation telescopes. We provided an example of these synergies in the cross-correlation between the 21cm signal and CII line intensity maps, where we could easily distinguish between galaxy properties consistent with different hydrodynamic simulations.

The simulation code is publicly available at \href{github.com/21cmFAST/21cmFAST}{github.com/21cmFAST/21cmFAST}, and can be used to explore the observability and constraining power of a range of observables including galaxy surveys, line intensity maps, and cosmic background radiation fields. The separation between the source field and the radiative background calculations will also allow outputs from other galaxy simulations and different radiative transfer algorithms to be inserted into our pipelines, allowing one to explore model dependence in both the galaxy models and the simulation algorithms.

\begin{acknowledgements}
      We thank G. Sun for helpful comments on a draft version of this manuscript.  We gratefully acknowledge computational resources of the HPC center at SNS.  AM acknowledges support from the Italian Ministry of Universities and Research (MUR) through the PRIN project "Optimal inference from radio images of the epoch of reionization", and the PNRR project "Centro Nazionale di Ricerca in High Performance Computing, Big Data e Quantum Computing".
\end{acknowledgements}

\bibliographystyle{aa}
\bibliography{paper}


\begin{appendix}
\begin{figure*}
	\includegraphics[width=\linewidth]{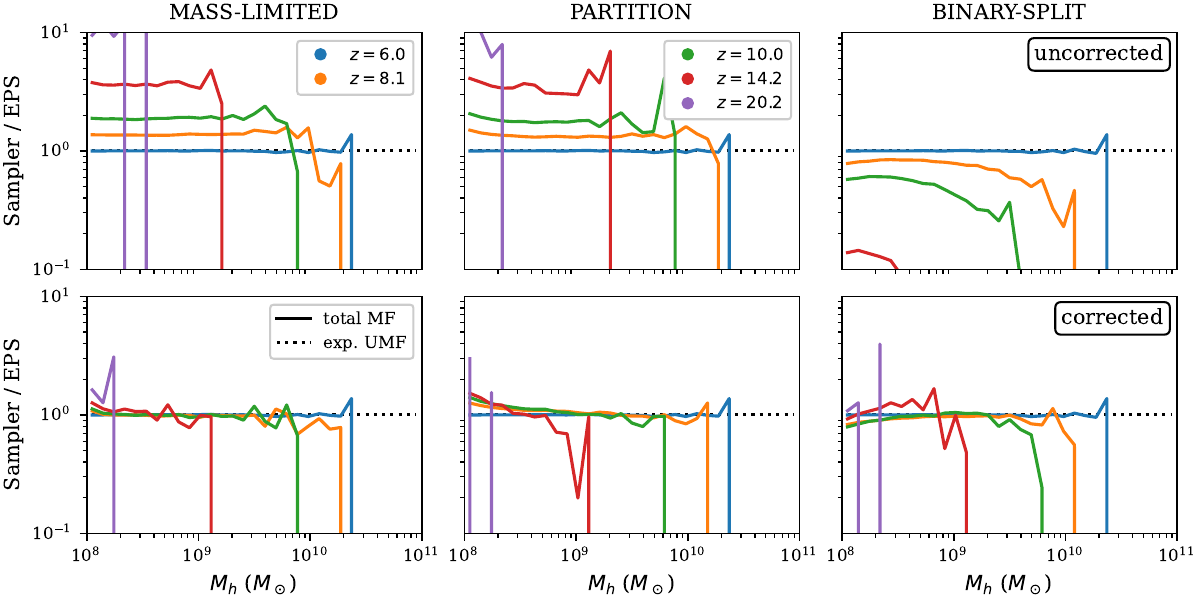}
    \caption{The mass functions of mean density $\delta=0$ Lagrangian cells utilising three different sampling methods. Left: the mass-limited sampling described in section \ref{sec:halosampling} Middle: The mass partitioning method described in \citet{Sheth99}. Right: The binary merger tree algorithm detailed in \citet{Parkinson08}. Top panels show the use of an unmodified EPS mass function in each case, bottom panels show adjustments made to the samplers (detailed in the text) in order to better match the expected EPS mass function.}
    \label{fig:cell_up}
\end{figure*}
\section{Comparison of sampling methods}\label{app:sampler}

In order to provide further justification of our method of sampling halo masses, we compare results here between our direct samples of the conditional mass function, described in section \ref{sec:sampling}, and two other methods for generating halo populations from a Lagrangian density distribution: the mass partitioning method from \citet{Sheth99}, where mass from a condition is allocated to halos according to the collapsed fraction distribution, and the binary merger tree algorithm from \citet{Parkinson08}, which uses small internal time-steps, taking the limit of the conditional mass function as the progenitor redshift approaches the descendant redshift $z_\mathrm{prog} \rightarrow z_\mathrm{desc}$. Both of these methods have been used for the efficient generation of halo populations \citep{McQuinn07,Benson16,Qiu21,Trinca22}, however their applications have so far been to produce single halo populations consistent with a certain N-body simulation, which has been achieved by fitting parameters in each model. The goal of our model is to produce halo populations consistent with \textit{any} given conditional mass function, and so directly sampling the given CHMF is a more natural choice.

We show mass functions in Lagrangian mean-density cells between redshifts $z=6$ and $z=30$ for each halo sampling method in Figure \ref{fig:cell_up}. All models shown utilise the Extended Press-Schechter CHMF, and the initial samples at $z=6$ are the same number-limited sampling described in section \ref{sec:halosampling}. We show two cases in the figure, the first using the EPS mass function with no corrective factors, and the second showing various corrective measures to each method so that the final halo samples are a better match the EPS mass function.

We see in the top panels of Figure \ref{fig:cell_up} that with no corrective factors, the \citet{Sheth99} partitioning method and our mass-limited sampling perform very similarly, with a slight excess at all progenitor masses which increases toward higher redshift. This excess originates from a lack of strict mass conservation in these models. Since we draw a discrete number of halos from the CHMF, we cannot guarantee that their masses sum to the expected mass above our resolution. Separate testing has shown that this excess is not strongly dependent on the choice of CHMF or the cell density $\delta$. This effect is partially mitigated in the mass-limited sampling case by utilising a strategy which has the opposite bias. Half of the time, once the mass limit is reached, we keep or throw away the final sampled halo depending on which decision brings us closer to the expected total mass. The other half of the time, we remove random halos from our sample until we are below the mass limit again, and then as before we keep the final halo selected for removal if it brings us closer to the expected mass. The result of this process is a $\sim 1\%$ excess in mass per timestep, which does not strongly depend on the halo mass, the condition, or the chosen CHMF.
The binary split algorithm on the other hand shows a defecit of halo mass compared to the EPS mass function, which increases both toward higher redshift and higher halo mass, consistent with the investigation performed in \citet{Cole08}.

In the bottom panels of Figure \ref{fig:cell_up} we demonstrate the effects of various corrective measures on these methods. For the mass-limited sampling we simply multiply the mass threshold for each halo by 0.9, which should be reflected as a downward shift in the HMF. For the partitioning algorithm we multiply each sampled peak $\nu = \frac{\delta - \delta_c}{\sqrt{\sigma - \sigma_c}}$ by 0.9, representing a leftward shift in the HMF. For the binary split algorithm we scale the EPS mass function by factors dependent on the condition mass and collapse threshold \citep{Parkinson08},

\begin{multline}
    n(M_h,z|\sigma_\mathrm{cond},\delta_\mathrm{cond}) =
    G_0 \left( \frac{\sigma(M_h)}{\sigma_\mathrm{cond}} \right)^{\gamma_1} 
    \left( \frac{\delta_\mathrm{cond}}{\sigma_\mathrm{cond}}\right)^{\gamma_2} \\
    n_\mathrm{EPS}(M_h,z|\sigma_\mathrm{cond},\delta_\mathrm{cond})
\end{multline}
where we use $G_0 = 1.0$ $\gamma_1 = 0.2$ $\gamma_2 = -0.3$ in order to roughly match the SMT conditional mass function. We see that in each case the sampled mass functions are much closer to the expected EPS mass function.

Since each sampling method is comparable in terms of speed, we make the decision to use the mass-limited sampling for this work as it is much more flexible for a general CHMF. The mass partitioning method gains a lot of its speed from the fast sampling of the gaussian EPS mass distribution. A more general CHMF slows the algorithm down due to the extra sampling steps required per halo sample, and it is not as amenable to tabulation of the CHMF since the condition changes throughout each sample, requiring an extra dimension in the tables.

It would be possible to fit the scaling factors in the binary split algorithm to any given CHMF, as has been done for N-body simulations \citep{Benson16,Qiu21}, and these scaling factors are provided as free parameters should a user wish to perform this analysis. However, since our mass limited sampler performs similarly in both speed and accuracy with the same corrective factor for both EPS and SMT conditional mass functions, we adopt this as our fiducial sampling method throughout this work.

\section{Code timing \& memory usage}\label{app:time}
\begin{figure}
    \centering
    \includegraphics[width=\linewidth]{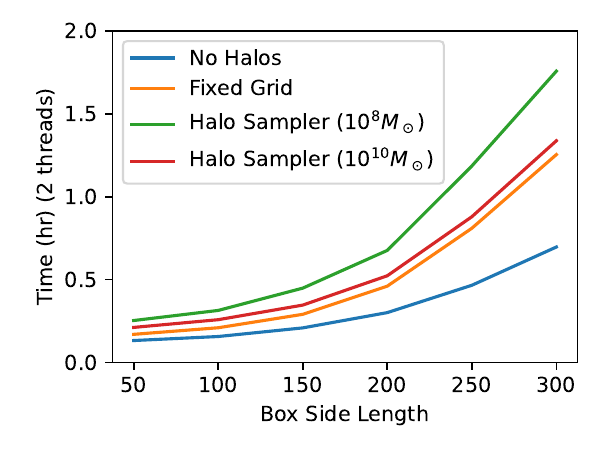}
    \caption{Time to run lightcones from $z=35$ to $z=6$ at various box sizes using 2 threads and a $2$ Mpc cell size. Blue: The previous source model of \cmfast (see ``v3 Default" runs in section \ref{sec:comparison} Orange: runs with a non-stochastic source field (see ``CHMF on grid" in section \ref{sec:comparison}). Green: Runtime for the halo sampler using a minimum mass of $10^8 M_\odot$. Red: Runtime for the halo sampler using a minimum mass of $10^{10}$. The new filters described in section \ref{sec:radiation} and the halo sampler at $10^8 M_\odot$ each add approximately $50\%$ to the runtime, compared with the grid-based model.}
    \label{fig:codetime}
\end{figure}

In figure \ref{fig:codetime} we show the total runtime of a lightcone using halo sampler, versus the fixed halo grids (``CHMF on Grid" in section \ref{sec:comparison}) and the default excursion set source model at various box sizes. All boxes run from $z=35$ to $z=6$ and use a cell size of $2$ cMpc. The halo sampler takes approximately double the runtime at all box sizes. The extra runtime is due in roughly equal parts to the halo sampler itself which performs calculations on a large number of halos, as well as the more complicated filter functions applied to the radiation fields (see section \ref{sec:radiation}). This can be seen in the difference between the halo sampling run and fixed halo grid run, which still uses the new filters, but does not need to sample any halos. To mitigate the increased requirements of the halo sampler, we provide an option which adds the average contribution from halos below the minimum sampled mass to the gridded source properties, sacrificing stochasticity in the smallest halo masses allowing the user to set a higher minimum mass. This speeds up the simulation and lowers memory requirements without losing the contribution of smaller sources. Setting the minimum mass to $10^{10} M_\odot$ improves the runtime significantly, approaching the time taken by the fixed halo grid run.

\begin{figure}
    \centering
    \includegraphics[width=\linewidth]{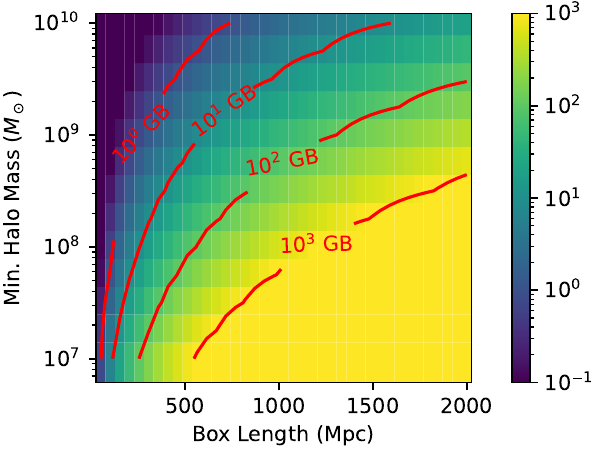}
    \caption{Memory usage of the halo sampler versus minimum sampled mass and box side length at redshift $z=6$. Contours are drawn at 1, 10, 100 and 1000 GB. Since the halo catalogs are the largest data objects in the simulation, this corresponds to the size of two halo catalogs (required for sampling from a descendant catalog) at $z=6$.}
    \label{fig:codemem}
\end{figure}

The halo sampler also requires more memory than the default \cmfast v3 run, and is strongly dependent on the minimum halo mass. We show the memory required by the halo sampler as a function of box size and minimum mass in figure \ref{fig:codemem}. The runs shown in this paper use a minimum mass of $10^8 M_\odot$ and a box size of $300 \mathrm{Mpc}$, which use $\sim 32 \mathrm{GB}$ of memory. Since every halo above the minimum mass is saved, the memory requirements are proportional to the integral of the unconditional mass function multiplied by the box volume. This is a significant increase in memory usage from the previous iteration when including stochastic sampling of small halos. However, since the average contribution of small halos are still included even when a higher minimum mass is used, a user without access to high-memory systems may still run relatively large, accurate simulations by setting the minimum mass to a larger value.

\end{appendix}

\end{document}